\documentclass[12pt]{iopart}

\pdfoutput = 1
\usepackage{bm}
\usepackage{color}
\usepackage{graphicx}
\usepackage{citesort}
\usepackage{cancel}
\eqnobysec

\newcommand{\gwig}{\mbox{\;\raisebox{.3ex}
    {$>$}$\!\!\!\!\!$\raisebox{-.9ex}{$\sim$}}\;}
\newcommand{\lambdabar}{{\hbox{$\lambda$\kern-1.ex\raise+0.45ex\hbox{--}}}}

\DeclareMathAlphabet{\mathpzc}{OT1}{pzc}{m}{it}

\begin{document}

\begin{flushright}
{\large \tt 
TTK-11-18}
\end{flushright}

\title[Bayes and present DM direct detection]{A Bayesian view of the current status of dark matter direct searches}

\author{Chiara~Arina\dag, Jan~Hamann\ddag\  and Yvonne~Y.~Y.~Wong\dag}

\address{\dag\  
Institut f\"ur Theoretische Teilchenphysik und Kosmologie, RWTH Aachen, 52056 Aachen, Germany}
\address{\ddag\  
Department of Physics and Astronomy, University of Aarhus, 8000 Aarhus C, Denmark}

\eads{\mailto{chiara.arina@physik.rwth-aachen.de}, \mailto{hamann@phys.au.dk}, \mailto{yvonne.wong@physik.rwth-aachen.de}}

\begin{abstract}
Bayesian statistical methods offer a simple and consistent framework for incorporating uncertainties into a multi-parameter inference problem. In this work we apply these methods to a selection of current direct dark matter searches. We consider the simplest scenario of spin-independent elastic WIMP scattering, and infer the WIMP mass and cross-section from the experimental data with the essential systematic uncertainties folded into the analysis. We find that when uncertainties in the scintillation efficiency of Xenon100  have been accounted for, the resulting exclusion limit is not sufficiently constraining to rule out the CoGeNT preferred parameter region, contrary to previous claims.  In the same vein, we also investigate the impact of astrophysical uncertainties on the preferred WIMP parameters. We find that within the class of smooth and  isotropic WIMP velocity distributions, it is difficult to reconcile the DAMA and the CoGeNT preferred regions by tweaking the astrophysics parameters alone. If we demand compatibility between these experiments, then the inference process naturally concludes that  a high value for the sodium quenching factor for DAMA is preferred.
\end{abstract}

\maketitle


\section{Introduction}

Recent years have seen a fervent activity in the direct search of Weakly Interacting Massive Particles (WIMPs) in the Galactic dark matter (DM) halo.
Besides the well-established results of DAMA/NaI and DAMA/Libra~\cite{Bernabei:2010mq}, which have observed altogether 13 successive cycles of annual modulation in the nuclear recoil rate consistent with the signature of Galactic WIMP scattering,
the CoGeNT experiment also claims an excess of events that cannot be accounted for by known background sources~\cite{Aalseth:2010vx}.
If interpreted as DM signals, then these results point to a particle mass of few GeVs in model independent analyses (e.g., ~\cite{Fitzpatrick:2010em,Chang:2010yk,Fitzpatrick:2010br,Hooper:2010uy,Schwetz:2010gv}), as well as in the frameworks of scalar DM (e.g., ~\cite{Andreas:2010dz,Barger:2010yn}), supersymmetric models (e.g.,~\cite{Fornengo:2010mk,Belikov:2010yi,Das:2010ww,Kappl:2010qx,Draper:2010ew}), and hidden sectors~(e.g., \cite{Kang:2010mh,Mambrini:2010dq}).  

Concurrently, the null results of several other direct detection experiments have led to exclusion limits in the WIMP parameter space.  For spin-independent scattering, CDMS~\cite{Akerib:2005kh,Ahmed:2009zw}, Xenon100~\cite{:2011hi}, Xenon10~\cite{Angle:2007uj}, Edelweiss~\cite{:2011cy}, the CRESST run on Tungsten~\cite{Angloher:2008jj}, and Zeplin-III~\cite{Lebedenko:2008gb} have all set relevant bounds. Most notably, the limits set by Xenon100 on the WIMP mass and cross-section appear to be incompatible with the regions preferred by 
the DM interpretation of the DAMA and CoGeNT results. Prudently though, we note that, depending on the detection techniques, direct WIMP searches can be subject to large 
systematic effects. Indeed, in the case of Xenon10, different choices of the scintillation efficiency ${\rm L_{eff}}$ can either enhance or reduce the compatibility between its exclusion limits and the DAMA/CoGeNT preferred parameters~\cite{Kopp:2009qt,Andreas:2010dz,Savage:2010tg}.

The first goal of this work, therefore, is to address the issue of how to account for systematic uncertainties in direct detection experiments. To this end, we employ the techniques of Bayesian inference. Bayesian methods provide a simple and consistent framework for dealing with  nuisance parameters---in this instance, poorly known
experimental parameters such as  ${\rm L_{eff}}$---in an inference problem. Once a likelihood function has been defined for an experimental result, the nuisance parameters can be systematically integrated out of the problem in a procedure known as marginalisation, 
yielding a final posterior probability density function (pdf) for the WIMP parameters that incorporates all relevant sources of uncertainties. Because the process of marginalisation requires the evaluation of a multi-dimensional integral, Markov Chain Monte Carlo (MCMC) methods
are particularly well-suited to the purpose. Lastly, we note that Bayesian inference is widely used for parameter estimation in precision cosmology (e.g.,~\cite{Trotta:2008qt}), and has recently also found application in high energy physics, e.g., for the exploration of supersymmetric parameter space (e.g.,~\cite{Baltz:2004aw,Allanach:2005kz,Trotta:2008bp}), or in view of forecasting model expectations for direct DM searches~\cite{Strigari:2009zb,Bertone:2010rv,Akrami:2010dn}.

In the same vein, the second goal of this work is to incorporate also into the picture some degree of uncertainty in the astrophysics. 
 The WIMP--nucleus scattering rate in a direct DM search depends on the (unknown) velocity distribution of the DM particles in the Galactic halo.  Because of its simplicity a common practice is to assume a Maxwellian distribution, 
in which the local DM density, the circular and the escape velocities are fixed at some ``standard'' values. However, these quantities are far from well-constrained by astrophysical observations.
Of even less certainty is our knowledge of the functional form of the WIMP velocity distribution.  Indeed,  simulations of structure formation suggest that substantial deviations from the isotropic Maxwellian form are highly probable (e.g.,~\cite{Vogelsberger:2008qb,Ling:2009eh,Kuhlen:2009vh,Peter:2009mi}). In this work we investigate several alternative velocity distributions.  For simplicity we consider only  isotropic equilibrium distributions consistent with selected spherically symmetric, smooth parametric DM halo density profiles motivated by $N$-body simulations.  Nevertheless, we see no obvious obstacle to generalising the analysis also to anisotropic velocity distributions (e.g.,~\cite{Ullio:2000bf,Belli:2002yt,Vergados:2007nc,MarchRussell:2008dy,Fairbairn:2008gz,Green:2010gw}) and non-smooth density profiles (e.g.,~\cite{Stiff:2001dq,Freese:2003na,Savage:2006qr,Green:2007rb,Schneider:2010jr}).

The rest of the paper is organised as follows:  after reviewing the basics of direct DM searches in section~\ref{sec:DDan}, we describe in section~\ref{sec:vel} various halo profiles and their corresponding WIMP velocity distributions. In section~\ref{sec:likeli} 
we construct the likelihood function for each experiment and  discuss the modeling of their associated systematics. Section~\ref{sec:stat} contains a detailed explanation of the Bayesian inference procedure. We present 
our inference results in section~\ref{sec:results}, and conclude in section~\ref{sec:concl}.


\section{The WIMP signal in direct detection experiments}\label{sec:DDan}

Direct detection experiments aim to detect or set limits on nuclear recoils arising from the scattering of WIMPs off target nuclei. 
The differential spectrum for such recoils, in units of  events per time per detector mass per energy, 
has the form
\begin{equation}
\label{eq:diffrate}
\frac{\rmd R}{\rmd E} = \frac{\rho_{\odot}}{m_{\rm DM} }  \int_{v'>v'_{\rm min}} \rmd^3v'  \, \frac{\rmd\sigma}{\rmd E} \, v'  \, f (\vec{v'}(t))\,,
\end{equation}
where $E$ is the energy transferred during the collision,  $\rho_{\odot} \equiv \rho_{\rm DM}(R_{\odot})$ the WIMP density in the solar neighbourhood,
$m_{\rm DM}$ the WIMP mass,
$\rmd \sigma/\rmd E$ the differential cross section for the scattering, and $f(\vec{v'}(t))$ is 
the WIMP velocity distribution in the Earth's rest frame normalised such that $\int \rmd^3 v' f(\vec{v'}(t))=1$. The integration in the differential rate is performed over all incident particles capable of depositing a recoil energy of $E$. For elastic scattering, this implies a lower integration limit of $v'_{\rm min} = \sqrt{M_{\cal N} E/2 \mu}$, where $M_{\cal N}$ is the mass of the target nucleus, and $\mu=m_{\rm DM} M_{\cal N}/(m_{\rm DM}+M_{\cal N})$ is the WIMP--nucleus reduced mass. We defer the discussion of the normalised velocity distribution $f(\vec{v'}(t))$ to section~\ref{sec:vel}.

The differential cross-section $\rmd \sigma/\rmd E$  encodes all the particle and nuclear physics information. For coherent elastic scattering it is parameterised as
\begin{equation}
\label{eq:pppart}
\frac{\rmd \sigma}{\rmd E} = \frac{M_{\cal N} \sigma^{\rm SI}_n}{2 \mu^2_n v'^2}\ \frac{\Big(f_p Z + (A-Z) f_n\Big)^2}{f_n^2} {\cal F}^2(E) \,  ,
\end{equation}
where $\mu_n=m_{\rm DM} m_n/(m_{\rm DM}+m_n)$ is the WIMP--nucleon
reduced mass, $\sigma^{\rm SI}_n$ the spin-independent (SI)
zero-momentum WIMP--nucleon cross-section, $Z$ ($A$) the atomic (mass)
number of the target nucleus used, and $f_p$ ($f_n$) is the WIMP
effective coherent coupling to the proton (neutron). 
We assume the WIMP couples equally to the neutron and the proton,  so that the differential cross-section $\rmd \sigma/\rmd E$ is sensitive only to $A^2$.  The nuclear form factor ${\cal F}(E)$ characterises the loss of coherence for nonzero momentum transfer, and in our analysis we use the Helm form factor~\cite{Helm:1956zz,Lewin:1995rx},
\begin{equation}
F(E)=3 e^{-k^2 s^2/2} \frac{\sin(kr)-kr \cos(kr)}{(kr)^3},
\end{equation}
where $s=1$~fm, $r=\sqrt{R^2-5 s^2}$, $R=1.2 \ A^{1/3}$~fm, and $k=\sqrt{2 M_{\cal N} E}$.

The total number of recoils expected  in a detector of mass $M_{\det}$ in a given {\it observed} energy range $[{\cal E}_1,{\cal E}_2]$ over an exposure time $T$ 
is obtained by integrating
equation~(\ref{eq:diffrate}) over energy,
\begin{equation}
\label{eq:totrate}
S(t)=M_{\rm det} T \int_{{\cal E}_1/q}^{{\cal E}_2/q}  \rmd E\  \epsilon(q E)\  \frac{\rmd R}{\rmd E} \,,
\end{equation}
where we have folded into the integral an energy-dependent function $\epsilon(q E)$ describing the efficiency of the detector.
The quenching factor $q$, defined via ${\cal E}=q E$, denotes the fraction of recoil energy that is ultimately observed in
a specific detection channel (scintillation or phonons/heat),  and is a detector-dependent quantity. To distinguish ${\cal E}$ from the 
actual nuclear recoil energy $E$, the former is usually given in units of keVee (electron equivalent keV), while the latter in keVnr (nuclear recoil keV).

\section{The WIMP velocity distribution}\label{sec:vel}

\subsection{Halo profiles\label{sec:haloprofiles}}

Two astrophysical factors enter into the differential recoil rate~(\ref{eq:diffrate}): the local WIMP density $\rho_\odot$ and the 
corresponding normalised velocity distribution $f(\vec{v'}(t))$ in the Earth's rest frame (primed $\vec{v'}$).  These quantities are related via
\begin{equation}
\rho_{\rm DM} (\vec{r}) = \int {\rm d}^3 v \  F(\vec{v},\vec{r})\, ,
\label{eq:Fv1}
\end{equation}
where $\rho_{\rm DM} (\vec{r})$ is the WIMP density at $\vec{r}$ from the Galactic Centre (GC) such that 
$\rho_\odot \equiv \rho_{\rm DM}(\vec{R}_\odot)$ with $R_\odot \equiv |\vec{R}_\odot|= 8.5$~kpc, and $F(\vec{v},\vec{r})$ is the WIMP velocity 
distribution in the Galactic frame (unprimed $\vec{v}$) whereby $f(\vec{v'}(t)) \equiv F(\vec{v},\vec{R}_\odot)/\rho_\odot$.

Most analyses in the literature assume a spherically symmetric and isothermal distribution for the WIMP around the GC.  The WIMP velocities follow the
Maxwellian distribution $F(\vec{v},r)\sim \exp(-v^2/\bar{v}^2)$, where $\bar{v}=220 \ {\rm km \ s}^{-1}$ is the mean velocity in the Galactic frame, 
and the distribution is cut off at $v > v_{\rm esc}=544 \ {\rm km \ s}^{-1}$.  The resulting density profile scales as $r^{-2}$~\cite{Binneybook}, and 
is normalised to $\rho_{\odot}=0.3\  {\rm GeV \ cm}^{-3}$.  This is known as the Standard Model Halo (SMH).

However,  isothermal DM density profiles are rarely if ever encountered in $N$-body simulations.  Indeed, most simulations find dark matter halos that are often ``cuspy'' and have density profiles that fall off faster than the $r^{-2}$ dependence of the SMH at large $r$.  In view of the uncertainty in the exact DM distribution, we consider, besides the SMH, four
other spherically symmetric DM density profiles found in the literature:
\begin{enumerate}
\item {\it Cored isothermal} \quad A variant of the SMH, this density profile has the form
\begin{equation}
\label{eq:isothermal}
  \rho_{\rm DM}(r) = \rho_s \, \left[ 1 + \left( \frac{r}{r_s}\right)^2 \right]^{-1}\,.
\end{equation}
Unlike the SMH in which the density $\rho_{\rm DM}(r)$ diverges as $r\to 0$, the cored isothermal halo has a finite density
core whose size and density are characterised by the parameters $r_s$ and $\rho_s$ respectively.  The profile's 
large $r$ (i.e., $r \gg r_s$) behaviour, however, is similarly to that of the SMH.

\item {\it Navarro--Frenk--White (NFW)} \quad Based on $N$-body simulation results, Navarro, Frenk and White suggested as a universal 
form for the DM density profile across a wide range of halo masses ($10^{11} \to 10^{15} M_\odot$)~\cite{Navarro:1996gj},
\begin{equation}
\label{eq:NFW}
  \rho_{\rm DM}(r) = \rho_s \, \left( \frac{r}{r_s} \right)^{- 1} \, \left( 1 + \left( \frac{r}{r_s}\right) \right)^{-2}\,.
\end{equation}
The density here falls off as $r^{-3}$ at $r \gg r_s$, while at $r \ll r_s$ we find a $r^{-1}$ behaviour (i.e., the cusp).
The NFW profile is formally divergent as $r\to 0$.  For numerical stability, however, we introduce in the profile a small core of size $\epsilon \ll r_s$. 
A related density profile is the Moore profile~\cite{Ghigna:1999sn}, which also exhibits a $r^{-3}$ behaviour at $r \gg r_s$, but has a steeper cusp that scales as $r^{-1.5}$.
We do not consider the Moore profile here because of its similarity to the NFW profile (see section~\ref{sec:veldist}).

\item {\it Einasto}  \quad Some recent studies find that the Einasto profile~\cite{1989A&A...223...89E} provides as good a fit as  the NFW profile to DM halos found in $N$-body simulations of the concordance $\Lambda$CDM model~\cite{Graham:2005xx}.
The Einasto profile has the form 
\begin{equation}
\label{eq:einasto}
  \rho_{\rm DM}(r)  =  \rho_s \exp\left(-\frac{2}{a} \left[\left(\frac{r}{r_s}\right)^a-1\right]\right) \,,
\end{equation}
where $a =0.17$, and its central density  is finite.

\item {\it Burkert} \quad The Burkert profile,
\begin{equation}
\label{eq:burkert}
  \rho_{\rm DM}(r) =  \rho_s \, \left( 1+ \frac{r}{r_s} \right)^{- 1} \, \left( 1 + \frac{r}{r_s} \right)^{-2}\,,
\end{equation}
is a cored profile that appears to provide a good fit to the DM distribution of dwarf galaxies~\cite{Burkert:1995yz}.
\end{enumerate}

All four profiles depend on two parameters $\rho_s$ and $r_s$.  However, it is equally valid, and perhaps more enlightening, to adopt a parameterisation in terms of 
the virial mass $M_{\rm vir}$ of the DM halo---defined as the mass contained in a sphere of radius $r_{\rm vir}$ whose average density is 200 times the critical density---and a concentration parameter given by $c_{\rm vir} = r_{\rm vir}/r_s$.  The advantage of this parameterisation is that, firstly, it is possible to specify directly 
a prior for $M_{\rm vir}$ based on what we know about the mass of the Milky Way from satellite kinematics etc.  Secondly, the concentration parameter $c_{\rm vir}$ is well studied in
$N$-body simulations, which again allow us to impose a prior on $c_{\rm vir}$ in a meaningful way.  We show how each density profile~(\ref{eq:isothermal}) to (\ref{eq:burkert})
can be expressed in terms of $M_{\rm vir}$ and $c_{\rm vir}$ in~\ref{app1}.

\subsection{Extracting the velocity distribution\label{sec:veldist}}

Given a DM density profile, the underlying DM velocity distribution can be extracted by inverting equation~(\ref{eq:Fv1}) under the assumption of 
hydrostatic equilibrium.
For a spherically symmetric density distribution 
and assuming an isotropic velocity distribution $F(\vec{v},r)=F(\varepsilon)$ in the Galactic frame that
depends only on the relative energy $\varepsilon \equiv \Psi - \frac{1}{2} v^2 \geq 0$ of the system, the solution is given by the Eddington formula~\cite{Binneybook},
\begin{equation}
\label{eq:eddie}
{ F(\varepsilon)} = \frac{1}{\sqrt{8}\pi^2}\left[ \int_{0}^{\varepsilon} \frac{\rmd^2\rho_{\rm DM}}{\rmd\Psi^2} \frac{\rmd\Psi}{\sqrt{\varepsilon - \Psi}} + \frac{1}{\sqrt{\varepsilon}} \left. \left(\frac{\rmd\rho_{\rm DM}}{\rmd\Psi}\right)
\right|_{\Psi=0}\right]\,.
\end{equation}
The function $\Psi(r)$ is the gravitational potential generated by the DM halo and the baryonic matter residing in the Galactic disk and bulge, 
defined so that $\Psi(r \to \infty) = 0$, and $v_{\rm esc}(r) \equiv \sqrt{ 2 \Psi(r)}$ is the escape velocity at $r$.   It is obtained by solving the 
Poisson equation, 
\begin{equation}
\label{eq:psi}
 \frac{\rmd^2 \Psi}{\rmd r^2} + \frac{2}{r} \frac{\rmd \Psi}{\rmd r} = -4 \pi G [\rho_{\rm DM}+\rho_{\rm disk}+\rho_{\rm bulge}],
\end{equation} 
where the disk density distribution is given by~\cite{Strigari:2009zb} 
\begin{equation}
  \label{eq:diskdiv}
  \rho_{\rm disk}(r) = \frac{ M_{\rm disk}}{4 \pi r_{\rm disk}^2} \, \frac{e^{-r/r_{\rm disk}}}{r},
\end{equation}
with $M_{\rm disk} = 5 \times 10^{10} \; M_\odot$ and $r_{\rm disk} = 4 {\rm \ kpc}$, and the bulge is modelled as a point mass sitting at $\vec{r}=0$, 
\begin{equation}
\rho_{\rm bulge}(r) = M_{\rm bulge} \delta^{(3)}_D(\vec{r}),
\end{equation}
where $M_{\rm bulge}=1.5 \times 10^{10} M_\odot$, and $\delta^{(3)}_D(\vec{r})$ is the 3-dimensional Dirac delta distribution.

At any given point $r$, the Eddington formula~(\ref{eq:eddie}) returns a positive and nonzero solution for $F(\varepsilon)$ only up to the escape velocity $v_{\rm esc}$ at that point.   For $v>v_{\rm esc}$, $F(\varepsilon)$ is by definition zero.  Furthermore, the formula shows that the DM velocity distribution at $R_\odot$, $F(\Psi_\odot-\frac{1}{2} v^2)$, depends only on the DM density distribution at $r>R_\odot$.  Thus, halo density profiles sharing the same large $r$ behaviour will yield similar solutions for  $F(\Psi_\odot-\frac{1}{2} v^2)$~\cite{Lisanti:2010qx}.   For this reason the NFW and the Moore profiles are for our purposes equivalent (see section~\ref{sec:haloprofiles}).

The last step is to rewrite the velocity integral in the differential recoil rate~(\ref{eq:diffrate}) in terms of  $F(\Psi_\odot-\frac{1}{2} v^2)$, that is,
\begin{equation}
\label{eq:fv2}
\int_{v'>v'_{\rm min}} \rmd^3v' \,  \frac{f (\vec{v'}(t))}{v'} \rightarrow 2 \pi \rho_\odot^{-1}\int_{v'>v'_{\rm min}}  \rmd v' \ v'  \int_{-1}^{1} \rmd \alpha \; F\Big(\Psi_{\odot}-\frac{1}{2} v^2 \Big)\,,
\end{equation}
with
\begin{eqnarray}
v^2 &=& | \vec{v'} + \vec{v}_\oplus |^2=v'^2 + v^2_\oplus + 2 v' v_\oplus \alpha\,, \nonumber \\
v_{\oplus} & = & |\vec{v}_\odot+\vec{v^{''}}_{\oplus,{\rm rot}} | = v_\odot  + v''_{\oplus,{\rm rot}}  \cos\gamma\cos [2 \pi (t-t_0)/T]\,,
\end{eqnarray}
where $\vec{v}_{\oplus}$  and $\vec{v}_\odot$ are, respectively,  the Earth's and the sun's velocity in the Galactic frame, 
$\vec{v''}_{\oplus,{\rm rot}}$ is the Earth's rotational velocity around the sun in the sun's rest frame, 
and $\gamma=60^{\circ}$ is the inclination of the Earth's rotation plane with respect 
the the Galactic plane.   For our analysis we take $v''_{\oplus,{\rm rot}} = 29.8 \ {\rm km} \ {\rm s}^{-1}$, and
$v_\odot = v_0 + 12 \ {\rm km} \ {\rm s}^{-1}$~\cite{Lewin:1995rx,Green:2003yh}, where
\begin{equation}
\label{eq:v0}
v_0\equiv 
 \left. \sqrt{- r \frac{{\rm d \Psi}}{{\rm d}r}} \right|_{r=R_\odot}
\end{equation}
is the circular velocity of the local standard of rest.
In the time-dependent piece the period $T$ is one year, while $t_0$ corresponds to June 2, the 
day on which $v_\oplus$ reaches its maximum.

Finally, we note that implicit in the  recoil rate~(\ref{eq:diffrate}) are four 
astrophysical observables: the local DM density $\rho_\odot$, the Milky Way virial mass $M_{\rm vir}$, 
the circular velocity of the local standard of rest $v_0$ defined in equation~(\ref{eq:v0}), and
the local escape velocity,
\begin{equation}
 v_{\rm esc} =   \left. \sqrt{2 \Psi}\right|_{r=R_\odot}\,.
\end{equation}
These can be independently constrained using observations of stellar and satellite kinematics.  We discuss this point in more detail in section~\ref{sec:astro}.

\section{Experiments and their likelihood functions}\label{sec:likeli}

The likelihood function ${\cal L}(X|\theta)$ denotes the probability of the data $X$ given some theoretical prediction $\theta$, 
 and plays a central role in Bayesian inference.   
In this section we describe the likelihood function used for each experiment, as well as the modelling of potential systematics. 
 An in-depth discussion of Bayesian methods is deferred to section~\ref{sec:stat}. 
Table~\ref{tab:prior1} summarises the free (MCMC) parameters of our analysis.

\subsection{CDMSSi}

The cryogenic CDMS experiment at the Soudan Underground Laboratory operates germanium and silicon solid-sate detectors.  Two events were 
observed at 55 and 95~keVnr in the silicon run (CDMSSi hereafter) on a $0.1$~kg detector in an exposure of 65.8 kg-days, compatible with an expected background of $B_n = 3.6$  neutrons and $B_e=0.8 \pm 0.6$  electrons 
in the $5\to100$~keVnr detection window~\cite{Akerib:2005kh}. No quenching factor is required for the CDMS experiment, i.e., $q=1$. 
For details of the detector efficiency $\epsilon(q E)$ we refer the reader to, e.g., \cite{Savage:2008er}.

We model the corresponding likelihood function with a Poisson(2) distribution,%
\footnote{The notation Poisson($n$) denotes the Poisson distribution 
for $n$ observed events.}
\begin{equation}
\label{eq:cdmssi}
\ln{\cal L}_{\rm CDMSSi} (2|S,B)= - S-B+2 + 2 \ln \left(\frac{S+B}{2} \right)\,,
\end{equation}
where $S$ is the expected WIMP signal in the detection window, and $B=B_n+B_e$ the expected background.
The likelihood function~(\ref{eq:cdmssi}) is normalised such that $\ln{\cal L}=0$ if the sum of the expected signal and background matches exactly
the number of observed events.

Since the expected background rate comes with an uncertainty---in this instance, $B=\bar{B}\pm \sigma_B=4.4 \pm 0.6$,
it is useful to construct an effective likelihood function  ${\cal L}^{\rm eff}$ by marginalising over
the background $B$,
\begin{equation}
\label{eq:bckm}
{\cal L}^{\rm eff}_{\rm CDMSSi} (2 |S) = \int_0^{\infty} \rmd B \ {\cal L}_{\rm CDMSSi}   (2 | S,B)\  p(B), 
\end{equation}
where 
\begin{equation}
\label{eq:bpdf}
p(B)  =\frac{1}{\sqrt{2 \pi \sigma_B^2}} \exp \left[-\frac{(B-\bar{B})^2}{2 \sigma_{B}^2} \right]
\end{equation}
is the probability density function of $B$  (modelled as a Gaussian distribution).
In the small $\sigma_B/\bar{B}$ limit, the resulting effective likelihood  has the form,
\begin{equation}
\label{eq:cdmssieff}
\ln{\cal L}_{\rm CDMSSi}^{\rm eff} = - S-\bar{B}+ \frac{\sigma_B^2}{2}+2 + \ln \left[\frac{ \sigma_B^2 + (S + \bar{B} - \sigma_B^2)^2}{4} \right] \, ,
\end{equation}
which we use in our inference analysis.%
\footnote{We adopt the small $\sigma_B/\bar{B}$ limit results whenever $\bar{B} \gwig 3 \sigma_B$.}

\begin{table}[t!]
\caption{MCMC parameters and priors for the WIMP parameter space and experimental systematics (nuisance parameters).  All priors
are uniform over the indicated range.\label{tab:prior1}}
\begin{center}
\lineup
\begin{tabular}{lll}
\br
Experiment& MCMC parameter & Prior \\
\mr
All & $\log(m_{\rm DM}/{\rm GeV})$  & $0  \to 3$\\
All & $\log(\sigma_n^{\rm SI}/{\rm cm}^2)$ & $-44 (-46) \to -38$\\ 
DAMA & $q_{\rm Na}$ &  $0.2 \to 0.4$\\
DAMA & $q_{\rm I}$ &  $0.06 \to  0.1$\\
Xenon100 & $m$& $-0.01 \to 0.18$ \\
CoGeNT & $C$ &  $0 \to 10$~cpd/kg/keV \\
CoGeNT & ${\cal E}_0$ &  $0 \to 30$~keV\\
CoGeNT & $G_n$ & $0 \to 10$~cpd/kg/keV\\
CDMSGe(LE) & $a$ & $-0.60 \to -0.18$\\
\br
\end{tabular}
\end{center}
\end{table}

\subsection{CDMSGe}

For their germanium run, the CDMS-II (CDMSGe hereafter) reported two events at 12.3 and 15.5~keVnr in the $10 \to 100$~keVnr window
in a total exposure of 612~kg-days~\cite{Ahmed:2009zw}.   The total expected background in the same time frame is $B= 0.8 \pm 0.1 \pm 0.2$.  For our analysis, however, we adopt the fitting formula for the differential background rate provided by~\cite{Kopp:2009qt}, 
\begin{equation}
\frac{\rmd N_B}{\rmd E} = \left[-0.00295 + 0.463 \left(\frac{\rm keVnr}{E} \right)\right]/(612~{\rm kg \ days})\,,
\end{equation}
where the rate has been normalised to $\bar{B}=0.8$ events over the detection window in an exposure of 612~kg-days.
Exploiting this spectral information, we model the likelihood function  as a product of two Poisson(1) distributions (for those energies with one event each) and a series
of Poisson(0) distributions (for those energies with no events)~\cite{loredo}, that is,
\begin{equation}
\label{eq:lcdmsge}
\ln{\cal L}_{\rm CDMSGe}  = -S-B+2 + \sum_{i=1,2}  \ln \left( \frac{\rm dR}{\rm dE_i} + \frac{B}{\bar{B}}\frac{\rmd N_B}{\rmd E_i} \right) +C_{\rm norm},
\end{equation}
where $S$ and $B$ are, respectively, the total expected signal and background in the detection window, $E_{1,2}=12.3,15.5$~keVnr, and
$C_{\rm norm}=\sum_{i=1,2}  \ln[M_{\rm det} T \epsilon(qE_i)]$  is a normalisation factor
following from the normalisation of the individual Poisson(1) and Poisson(0) distributions.  See discussion after equation~(\ref{eq:cdmssi}).

Marginalising over the total background $B$ (but not the spectral shape)
 in the manner of
 equations~(\ref{eq:bckm}) and (\ref{eq:bpdf}), we find in the small $\sigma_B/\bar{B}$ limit an effective likelihood 
\begin{eqnarray}
\label{eq:cdmsgeeff}
\ln{\cal L}_{\rm CDMSGe}^{\rm eff} &=& - S-\bar{B}+ \frac{\sigma_B^2}{2}+2 +C_{\rm norm}+\nonumber \\
&&  \ln \left[\prod_{i=1,2} \left(\frac{\rmd R}{\rmd E_i} + \frac{\bar{B}-\sigma^2_B}{\bar{B}} \frac{\rmd N_B}{\rmd E_i} \right) 
+\sigma_B^2 \prod_{i=1,2} \frac{1}{\bar{B}} \frac{\rmd N_B}{\rmd E_i} \right].
\end{eqnarray}
We include in the analysis also  null results from three previous searches with the CDMS germanium detector, 
with exposures of 34~kg-days~\cite{Akerib:2005kh}, 19.4~kg-days~\cite{Akerib:2004fq},  and  
397.8~kg-days~\cite{Ahmed:2008eu}, bringing the total exposure to 1063.2~kg-days.  The expected background and its
uncertainty are scaled correspondingly to $\bar{B}= 1.39$ and $\sigma_B=0.38$ respectively. We model
the detector efficiency $\epsilon(q E)$ after~\cite{Kopp:2009qt}.

\paragraph{Low energy CDMS}
The CDMS collaboration has recently re-analysed their germanium data---both on their own, and in combination with data from the silicon detectors---with a lower energy threshold~\cite{Ahmed:2010wy,Akerib:2010pv}, thereby increasing the experiment's sensitivity to light WIMPs.  
In reference~\cite{Ahmed:2010wy}, data from 8 germanium detectors (CDMSGe(LE) hereafter) were re-analysed using a threshold of 2~keVnr, compared to 10~keVnr in the standard analysis. For each detector, the collaboration provides the event energies and the raw exposure.  After summing up all  contributions and applying the efficiency cuts one finds a total of 427 
counts for 214 kg-days, distributed in the energy range $2 \to 100$~keVnr. In our analysis, we bin the data in such a way that 16 bins are contained in the energy 
range $2 \to 10$~keVnr, and 9 in $10 \to 100$~keVnr.

A lower energy threshold, unfortunately, is traded at the cost of an increased acceptance of background events, because at these low energies the ability of the experiment to 
discriminate between nuclear and electron recoils degrades and the ionization signal becomes dominated by noise. Indeed, while the
 background due to surface events, ``zero-charge'' events, and leakage events are reasonably well known at energies $>5$~keVnr, 
 between 2~keVnr and 5~keVnr the CDMS collaboration has to rely on an extrapolation to model these events in their analysis, 
 as described in figure~1 of~\cite{Ahmed:2010wy}. 
 Another potential issue is the calibration of the recoil energy near threshold, since the ionisation signal is missing.

Given these considerations, we model the differential background rate in our analysis as
\begin{equation}
\label{eq:mb}
{\rm m}_{\rm B} (E)= \left\{ \begin{array}{ll} \bar{\rm m}_{\rm B}(E), & \quad E \geq 5~{\rm keVnr},  \\
                                            0.1 \times 10^{a [(E/{\rm keVnr}) -5]}, & \quad 2<E/{\rm keVnr} <5 ,    \end{array} \right.
\end{equation}
where $ \bar{\rm m}_{\rm B}(E)$ corresponds to the black curve in figure~1 of~\cite{Ahmed:2010wy}, which, for energies above 5~keVnr,   
we regard as reliable and free of systematics. For energies below 5~keVnr we use an extrapolation function, but allow the slope $a$ to vary
subject to a Gaussian constraint
\begin{equation}
\ln {\cal L}_{m_{\rm B}} = - \frac{(a-\bar{a})^2}{2 \sigma_a^2},
\end{equation}
where the ``best-fit'' $\bar{a} = -0.36$ reproduces the black curve in figure~1 of~\cite{Ahmed:2010wy}, and 
$\sigma_a = 0.2 \ \bar{a}$, chosen based on the error bars in the same figure at 5 keV.

The expected signal rate in the $i$th energy  bin is then a sum of the DM signal and the background rate,
\begin{equation}
s_i = \frac{1}{\Delta E_{i}}  \int_{E_{i}-\Delta E_i/2}^{E_i+\Delta E_i/2}\!  \rmd  E\  \left[ \frac{{\rmd}R}{{\rmd}E} 
 + {\rm m_{B}}(E) \right],
\end{equation}
where $\Delta E_i$. The likelihood function  is given by
\begin{equation}
\label{eq:cdmsgele}
 \ln\mathcal{L}_{\rm CDMSGe(LE)} = - \sum_{i=1}^{N_{\rm bin}} \frac{(s_i-\bar{s}^{\rm  obs}_i)^2}{2 \sigma_i^2} + \ln\mathcal{L}_{m_{\rm B}} \,. 
\end{equation}
where $\bar{s}^{\rm  obs}_i$ is the observed rate in the $i$th bin, and $ \sigma_i$ is the associated error.

We do not consider the re-analysis of the combined data on the germanium and the silicon towers presented in~\cite{Akerib:2010pv}, because the lack of knowledge about the low-energy background makes it difficult for us to model the likelihood function.

\subsection{CoGeNT}  

The CoGeNT experiment, an ultra low-noise (and hence  low-threshold: 0.4~keVee) germanium cryogenic detector running at the Soudan Mine,
found in a total exposure of 18.48~kg-days an excess at low energies that cannot be attributed to known background sources~\cite{Aalseth:2010vx}.
Using the energy binning in figure~3 of~\cite{Aalseth:2010vx} in the $0.4 \to 3.2$~keVee energy range, we model the likelihood function as a sum of Poisson($X_i$) distributions,
\begin{equation}
\label{eq:cogent}
\ln{\cal L}_{\rm CG} = \sum_{i=1}^{56} \left[ -s_i-b_i-r_i+X_i +X_i \ln \left(\frac{s_i+b_i+r_i}{X_i}\right)\right],
\end{equation}
where $X_i$ is the number of events observed in the $i$th energy bin, $s_i$ is the expected signal computed
from equation~(\ref{eq:totrate}) with ${\cal E}_1$ and ${\cal E}_2$ corresponding respectively to the lower and upper energy limits of the bin concerned,
and $b_i$ and $r_i$ are two background components.

For the first component $b_i$, we model the differential rate as an exponentially decaying function,
\begin{equation}
\frac{\rmd N_b}{\rmd {\cal E}} = C \exp(-{\cal E}/{\cal E}_0),
\end{equation}
where $C$ and ${\cal E}_0$ are two free parameters.  
The second component $r_i$ denotes events due to two radiation peaks from $^{65}$Zn and $^{68}$Ge decays, whose differential rates are modelled as
Gaussians centered on the energies ${\cal E}_{\rm Zn}=1.1\  \rm keVee$ and ${\cal E}_{\rm Ge}=1.29 \ \rm keVee$, with a common standard deviation fixed by the energy resolution of the detector 
$\Delta {\cal E}$, i.e.,
\begin{equation}
\frac{\rmd N_{\rm Zn}}{\rmd {\cal E}}=G_{n,{\rm Zn}} \exp\left(-\frac{({\cal E}-{\cal E}_1)^2}{2 \Delta {\cal E}^2} \right),
\end{equation} 
and similarly for $dN_{\rm Ge}/dE$. We fix the ratio of the two peak heights to $G_{n,{\rm Zn}}/G_{n,{\rm Ge}}=0.7$, and vary only $G_n \equiv G_{n,{\rm Ge}}$. For details about cosmogenic backgrounds we refer to~\cite{phdthesis}. The energy resolution is given by $\Delta {\cal E}/{\rm eVee} = \sqrt{\sigma_n^2 +2.96\ F \ ( {\cal E}/{\rm eVee}) }$, with $\sigma_n=69.4$ and $F=0.29$~\cite{Aalseth:2010vx}, while the energy-dependent quenching factor is taken to be 
$q=2/[1+\sqrt{1+15.55 \ ({\rm keVee}/{\cal E}})]$, following~\cite{Kopp:2009qt}.

Because the background model parameters  $C$, ${\cal E}_0$ and $G_n$ enter into the likelihood function~(\ref{eq:cogent}) in a nontrivial fashion, analytical marginalisation 
in the manner of equation~(\ref{eq:bckm}) is cumbersome if not impossible.  We therefore treat these parameters as MCMC parameters.

\subsection{Xenon}\label{sec:xenon}

Xenon is a two phase (liquid/gas) xenon experiment running at Laboratori Nazionali del Gran Sasso (LNGS).  A nuclear recoil from particle scattering 
is inferred from the simultaneous measurements of scintillation light  and ionisation electrons, together with the arrival direction.
The amount of nuclear recoil energy going into the primary scintillation signal is expressed in terms of the number of photoelectrons (PE) produced $S_1$, which is related to the nuclear recoil energy $E$  through the relation

\begin{equation}
\label{eq:s1leff}
S_1 (E)= {\rm L_{\rm eff}}(E)\ {\rm L}_y\  E\ \frac{S_{\rm nr}}{S_{\rm ee}} \, ,
\end{equation}
where ${\rm L_{eff}}(E)$ is the  energy-dependent scintillation efficiency, ${\rm L}_y = 2.2 \ {\rm PE/keVee}$ the scintillation efficiency of nuclear recoils relative to that of the 122~keVee $\gamma$-rays at zero field, and the quantities $S_{\rm nr,ee} = 0.95,0.58$ denote respectively the electric field scintillation quenching factors for nuclear and electron recoils.

Between 2006 and 2007, the Xenon10 collaboration found 13 events for an expected background of 7 events in an exposure of 316.4 kg-days in the $2.0 \to 75.0$~keVnr window~\cite{Angle:2007uj,Angle:2009xb}.
The Xenon100 experiment recently released an analysis of 100.9 live days of data acquired in 2010, which found three candidate events
for an expected background of $B = 1.8 \pm 0.6$ in an exposure of 1481 kg days.~\cite{:2011hi}.  Because of this large exposure,
we consider in this work only the results of Xenon100 from the aforementioned data release, although the analysis techniques can easily be generalised for used with Xenon10.

As in the case of CDMSGe we include both the total rate and spectral information in the likelihood,
\begin{equation}
\label{eq:xe100}
\ln{\cal L}_{\rm Events} = - S - B + 3 + \sum_{i=1}^3 \ln \left(\left.\frac{\rmd R}{\rmd S_1}\right|_i + \frac{B}{\bar{B}} \left.\frac{\rmd N_B}{\rmd S_1}\right|_i \right) + C_{\rm norm}\,,
\end{equation}
where three events are at 8, 20, and 23 PE, respectively, and $C_{\rm norm}=\sum_{i=1,2,3} \ln( M_{\rm det} T)$ is a constant normalisation factor.
The background has been shown to consist mainly of electron recoils, with a flat distribution in energy over the experimental range
according to measurements and Monte Carlo simulations~\cite{Aprile:2011hx}.
We therefore model the differential background as 
$\rmd N_B/\rmd S_1 = 0.069/(1481 \ {\rm kg} \ {\rm days})$, normalised to $\bar{B}=1.8$ in the detection window of $4 \to 30$~PE in 1481 kg days.

The expected WIMP signal $S$  is computed as follows.
Firstly, we note that the actual conversion between the nuclear recoil energy and the number of PEs
produced is not deterministic. Given some recoil energy $E$, the number of PEs produced $S_1$ 
is subject to Poisson fluctuations, with 
equation~(\ref{eq:s1leff}) expressing only the expectation value $\bar{S}_1$.   In physical terms this means
recoils below the nominal energy threshold have a finite probability of leaking into the detection window
of the experiment, and this effect is important for the detection of light WIMPs.
The expected number of WIMP events as a function of the (discrete) number of PEs generated can be written as
\begin{equation}
\label{eq:Pconv}
\frac{\rmd R}{\rmd S_1} = \int_{0}^{\infty} \rmd E\  \frac{\rmd R}{\rmd E} \times P(S_1|\bar{S}_1(E))\,,
\end{equation}
where $P(S_1|\bar{S}_1(E))$ denotes a Poisson($n$) distribution with expectation $\bar{S}_1(E)$.
Summing over all PE counts, the total number of events expected in the detector is
\begin{equation}
S = M_{\rm det} T \sum_{n = {\rm PE_{min}}}^{\rm PE_{max}} \frac{\rmd R}{\rmd S_1}\,,
\end{equation}
where for Xenon100, ${\rm PE_{min}}=4$ and ${\rm PE_{max}}=30$.

It remains to specify the energy-dependent scintillation efficiency ${\rm L_{\rm eff}}(E)$.  In this work we use,
\begin{eqnarray}
\label{eq:leff}
{\rm L}_{\rm eff} (E) = \left\{ \begin{array}{ll} \bar{\rm L}_{\rm eff}(E), & \, E \geq 3~{\rm keVnr},  \\
                                             {\rm max}\{m [\ln(E/{\rm keVnr})\! -\! \ln3] \!+ \! 0.09,\ 0\}, & \, 1<E/{\rm keVnr} <3.     \end{array} \right. \nonumber \\
\end{eqnarray}
Here, $\bar{\rm L}_{\rm eff}(E)$ corresponds to the best-fit in figure~1 of~\cite{:2011hi}, which, at 
$E \geq 3$~keVnr, is well-constrained by direct measurements.  No direct measurements exists at $1<E/{\rm keVnr}<3$, and 
the ``best-fit'' provided by the Xenon100 collaboration is merely an extrapolation.  We therefore treat the extrapolation 
slope $m$ as a variable, MCMC parameter, subject to a Gaussian constraint of 
\begin{equation}
\ln {\cal L}_{\rm L_{\rm eff}} = - \frac{(m-\bar{m})^2}{2 \sigma_m^2},
\end{equation}
where $\bar{m}\equiv 0.082$ reproduces the ``best-fit'' of~\cite{:2011hi}
in the $1<E/{\rm keVnr}<3$ region, and $\sigma_m=0.04$, chosen so that the $2\sigma$ region
coincides approximately with the light blue band.  We further restrict $m$ to lie within the range $[-0.01,0.18]$,
so that $L_{\rm eff}(E)$ never exceeds $0.1$ at $E=1$~keVnr, or drops to zero at energies above 2~keVnr.

We note that a somewhat different parameterisation for the uncertainty in ${\rm L_{\rm eff}}(E)$
was adopted in~\cite{Aprile:2011hx}, which also accounts for errors in the direct measurements of 
${\rm L_{\rm eff}}(E)$ at $E \geq 3$~keVnr.
Nonetheless we choose the 
parameterisation~(\ref{eq:leff}) because it highlights the role of ${\rm L_{\rm eff}}(E)$ in the low energy region. 
Indeed, from equation~(\ref{eq:Pconv}), we see that the main uncertainty in the expected event rate at energies close to the threshold 
comes from  Poisson fluctuations in the number of 
PEs produced, which in turn depend sensitively on the unknown slope of ${\rm L}_{\rm eff}(E)$ at $E<3$~keVnr via equation~(\ref{eq:s1leff}). The small errors around the best-fit ${\rm L}_{\rm eff}(E)$ at  $E>3$~keVnr, on the other hand, has little impact on the physics close to the threshold and hence also the exclusion power of the experiment for light WIMPs. Note that our likelihood function~(\ref{eq:xe100}) automatically takes into account uncertainties in $L_{\rm eff}(E)$ in both the total number of signal events and their energy dependence,
in contrast to the approach of~\cite{Aprile:2011hx}, which explicitly assumes the (normalised) signal energy spectrum to be independent of these uncertainties.

Thus, the full likelihood function describing the Xenon100 experiment is
\begin{equation}
\ln{\cal L}_{\rm Xenon}  = \ln{\cal L}_{\rm Events} + \ln{\cal L}_{\rm L_{eff}}\,,
\label{eq:lkhxetot}
\end{equation}
which we further marginalise numerically over the background events $B$ (in the $\ln{\cal L}_{\rm Events}$ term) as per equations~(\ref{eq:bckm}) and (\ref{eq:bpdf}),
yielding an effective likelihood that depends only on $m_{\rm DM}$, $\sigma^{\rm SI}_n$, and the systematics nuisance parameter $m$.

The Xenon10 collaboration recently published a low-energy analysis
based on the ionisation signal (the $S_2$ signal)~\cite{Angle:2011th}.
This alternative approach removes the dependence of the result on the
uncertainties of the scintillation efficiency, thereby lowering the
detector threshold significantly down to 1~keV.  However, without a
reliable parameterisation of the estimated background, it is not
possible to construct a meaningful likelihood function for our
Bayesian analysis.  Since the Xenon10 collaboration does not provide
the necessary information, we refrain from using their data in this
work.

\subsection{DAMA}\label{sec:dama}

The DAMA/Libra~\cite{Bernabei:2008yi,Bernabei:2010mq} experiment at LNGS  uses NaI(Tl) crystal radio-pure scintillators as targets. The signature of  WIMP interactions consists of  an annual modulation of the signal due to the motion of the Earth through the Galactic halo  as discussed in section~\ref{sec:veldist}. 
For a cumulative exposure of 1.17 ton-year, the collaboration reported a positive detection at $8.9\sigma$ significance.

For  isotropic WIMP velocity distributions  such as those considered in this work, the expected modulation signal
averaged over an observed energy interval $[{\cal E}_1,{\cal E}_2]$  is given by
\begin{equation}
\label{eq:sm2}
s = \frac{1}{{\cal E}_2-{\cal E}_1} \!  \sum_{X={\rm Na},{\rm I}} w_X \!
\int_{{\cal E}_1/q_X}^{{\cal E}_2/q_X }\!  \rmd  E\   \frac{1}{2} \left[\frac{{\rmd}R_X}{{\rmd}E} 
(\mbox{\rm June} \ 2)-
 \frac{{\rmd}R_X}{{\rmd}E} (\mbox{\rm Dec} \ 2)  \right] ,
\end{equation}
where  $w_X \equiv M_X/(M_{\rm Na}+M_{\rm I})$,  and we have explicitly ignored the small contribution from channelling~\cite{Bozorgnia:2010xy}.
The likelihood follows a Gaussian distribution,
\begin{equation}
 \ln\mathcal{L}_{\rm DAMA} = - \sum_{i=1}^{N_{\rm bin}} \frac{(s_i-\bar{s}^{\rm  obs}_i)^2}{2 \sigma_i^2} \,, 
\end{equation}
where $s_i$ and $\bar{s}^{\rm obs }_i$ are the theoretical and the mean observed modulation respectively in the $i$th energy bin, $\sigma_i$ is the associated 
uncertainty in the observed signal, and we use in this analysis  the 36-bin data from figure~9 of~\cite{Bernabei:2008yi}. The quenching factors $q_{\rm Na}$ and $q_{\rm I}$ are taken to be free parameters in our analysis, which 
we vary, respectively, over a range representative of the diverse measured values found in the 
literature~\cite{Bernabei:1996vj,Chagani:2008in,Fushimi:1993nq,Smith:1996fu}.

\begin{table}[t!]
\caption{Additional MCMC parameters and (uniform) priors related to the modelling of the WIMP velocity distribution.\label{tab:prior2}}
\begin{center}
\lineup
\begin{tabular}{lll}
\br
Density profile & MCMC parameter & Prior\\
\mr
All & $M_{\rm vir}$  & $1 \to 5 \times 10^{12} \ M_\odot$ \\
NFW, Einasto & $c_{\rm vir}$ &  $5  \to 20$~\cite{Navarro:2008kc}\\
Cored isothermal, Burkert &  $c_{\rm vir}$ & $50 \to 200$~\cite{Jimenez:2002vy}\\
 \br
\end{tabular}
\end{center}
\end{table}

\subsection{Other experiments}\label{sec:exp}

Even though we will not consider their results in our analysis, let us
also mention the following direct detection experiments.

The CRESST collaboration has found 32 events on oxygen given an
expected background of $8.7\pm1.4$~\cite{CRESST}.  If interpreted as a
WIMP signal, this would point to a low mass WIMP with a coherent
cross-section in the ballpark of the regions preferred by DAMA and
CoGeNT data~\cite{Schwetz:2010gv}.

The Edelweiss collaboration recently published the final analysis for
their second run, reporting 5 events for an expected background of 3,
4 of which are close to the threshold~\cite{:2011cy}. 
An exclusion bound was set, which, because of the smaller exposure of
the experiment, is not competitive with those derived from other
experiments considered in this work. A recent combined analysis of
Edelweiss and CDMS has improved the sensitivity: a tighter exclusion
bound for DM masses above 200~GeV was found, relative to the limits
obtained by the individual experiment alone~\cite{CDMS:2011gh}.
However, this new limit is still less constraining than that derived
from the Xenon100 experiment.

\begin{table}[t!]
\caption{Astrophysical constraints on the DM halo profile and the WIMP velocity distribution.\label{tab:prior3}}
\begin{center}
\lineup
\begin{tabular}{ll}
\br
Observable & Constraint  \\
\mr
Local standard of rest&  $v_0^{\rm obs} = 230 \pm 24.4 \ {\rm km \ s}^{-1}$~\cite{Reid:2009nj,Gillessen:2008qv}\\
Escape velocity &   $v_{\rm esc}^{\rm obs}= 544   \pm 39 \  {\rm km \ s}^{-1}$~\cite{Smith:2006ym,Dehnen:1997cq} \\
Local DM density & $\rho_{\odot}^{\rm obs} = 0.4 \pm 0.2 \  {\rm GeV \ cm}^{-3}$~\cite{Weber:2009pt,Salucci:2010qr} \\
Virial mass &  $M_{\rm vir}^{\rm obs} = 2.7  \pm 0.3 \times 10^{12} M_{\odot}$~\cite{Dehnen:2006cm,Sakamoto:2002zr} \\
\br
\end{tabular}
\end{center}
\end{table}

\subsection{Astrophysics\label{sec:astro}}

In addition to the WIMP mass, cross-section, and the nuisance parameters of the direct search experiments, two further free parameters are used to characterise 
the WIMP velocity distribution: the virial mass of the DM halo, and its concentration (see table~\ref{tab:prior2}).  These additional parameters are, however, also constrained
by astrophysical observations.  For this reason, we define a likelihood function for the astrophysics,
\begin{eqnarray} 
\label{eq:lkhastro}
\ln{\cal L}_{\rm Astro} =\! -  \frac{(v_0 - \bar{v}^{\rm obs}_0)^2}{2 \sigma^2_{v_0}} \! -  \! \frac{(v_{\rm esc} - \bar{v}^{\rm obs}_{\rm esc})^2}{2 \sigma^2_{v_{\rm esc}}} \!- \! \frac{(\rho_\odot - \bar{\rho}^{\rm obs}_\odot)^2}{2 \sigma^2_{\rho_\odot}}\!-  \! \frac{(M_{\rm vir} - \bar{M}^{\rm obs}_{\rm vir})^2}{2 \sigma^2_{M_{\rm vir}}}, \nonumber\\
\end{eqnarray}
where the measured values of the various astrophysical observables
(see section~\ref{sec:veldist} for their definitions) and their
uncertainties are given in table~\ref{tab:prior3}.  Note that none of
the constraints in table~\ref{tab:prior3} assumes a specific
parameterisation of the halo profile, which allows us to apply them to
all halo models we are considering here without running the risk of
double-fitting.


\section{Statistical inference}\label{sec:stat}

Having specified a theoretical model with free parameters $\theta$ in
sections~\ref{sec:DDan} and \ref{sec:vel}, and defined the likelihood
functions $\mathcal{L}(X|\theta)$ in section~\ref{sec:likeli}, one final
step remains to be taken in the analysis of the data $X$: the
inference of the posterior probability density as a function of the
parameters, $\mathcal{P} (\theta | X)$.  
The posterior pdf
represents our state of knowledge about the parameters after taking
into account the information contained in the data, and has an
intuitive and straightforward interpretation in that $\int_V
\mathcal{P} (\theta | X) {\rm d}\theta$ is the probability that the
true value of $\theta$ lies in the volume $V$.    Given a likelihood function, 
the posterior pdf can be constructed
by invoking Bayes' theorem,
\begin{equation}
  \mathcal{P} (\theta | X) {\rm d}\theta \propto\ \mathcal{L}(X |
  \theta) \cdot  \pi(\theta) {\rm d}\theta \,,
\end{equation}
but the construction requires us to specify $\pi(\theta) $, the
probability density on the parameter space $\theta$ prior to observing the data
$X$.   Since this prior pdf is independent of the data, it needs to be
chosen according to one's theoretical prejudice, and is thus inherently subjective.

In the often encountered situation in which no unique theoretically
motivated prior pdf can be derived, one may wish to use one which
does not favour any parameter region in particular.
A common choice in this
case is the top-hat, or uniform, prior
\begin{equation}
\pi_{\rm flat}(\theta)  \rmd \theta \propto \left\{ 
    \begin{array}{cl} 
      \rmd \theta , &
     {\rm if}\ \theta_{\rm min} \leq \theta \leq  \theta_{\rm max} ,
     \\ 0, & {\rm otherwise},
    \end{array}
\right.
\end{equation}
if the general order of magnitude of the parameter is known.   Here, the limits $\theta_{\rm min}$ and $\theta_{\rm max}$ should 
be chosen such that they are well beyond the parameter region of interest. If even
the order of magnitude is unknown, one may want to choose a uniform prior in $\log \theta$ space 
instead,
\begin{equation}
\pi_{\rm log}(\log \theta) \ \rmd \log \theta  = \left\{ 
    \begin{array}{cl} 
   \rmd \log \theta, &
      {\rm if} \ \theta_{\rm min} \leq  \theta \leq  \theta_{\rm max},
      \\ 0, & {\rm otherwise},
    \end{array}
\right.
\end{equation}
which 
is equivalent to a $\rmd \theta/\theta$ prior in $\theta$ space. Note that
because the volume element $ \rmd \theta$ is in general not invariant under a parameter
transformation $f: \theta \to \theta'$, a uniform prior pdf on $\theta$ does not yield the same 
probabilities as a uniform prior pdf on $\theta'$ unless the mapping $f$ is linear.  
The same is also true for the posterior probabilities, i.e., $  \mathcal{P} (\theta | X) {\rm d}\theta \neq
  \mathcal{P} (\theta' | X) {\rm d}\theta'$ in general.

While the posterior pdf technically contains all the necessary information for the 
interpretation of  the data, the fact that it is a function in the
$N$-dimensional space of parameters makes it difficult to
visualise if $N>2$.  Fortunately, by virtue of being a probability
density, its dimensionality can be easily reduced by
integrating out less interesting ({\it nuisance}) parameter directions
$\psi_i$, yielding an $n$-dimensional marginal posterior pdf,
\begin{equation}
 \label{eq:marg}
 \mathcal{P}_{\rm mar}(\theta_1, ..., \theta_n | X) \propto \int \rmd\psi_1
 ... \rmd\psi_m \ {\cal P}( \theta_1, ..., \theta_n,\psi_1...,
 \psi_m|X) \,, 
\end{equation} 
which is more amenable to visual presentation if $n=1,2$, and can
be used to construct constraints on the remaining parameters.

A complementary approach to the marginalisation 
is to project the likelihood function $\mathcal{L}(X|\theta)$ onto the $n$-dimensional subspace by maximising along the nuisance directions, i.e,
\begin{equation}
 \label{eq:profile}
 \mathcal{L}_{\rm prof}(X|\theta_1, ..., \theta_n ) \propto  \max_{\psi_1
 ... \psi_m} \ {\cal L}( X|\theta_1, ..., \theta_n,\psi_1...,
 \psi_m) \,.
\end{equation}
Maximisation is not a Bayesian procedure, and the resulting  profile likelihood
cannot be interpreted as a probability density function.  However, because 
$\mathcal{L}_{\rm prof}$ is by construction insensitive to our choice of priors and associated 
volume effects, it can be a useful 
means to assess if the inference has been significantly affected by our choice of nuisance parameterisation.%
\footnote{We always normalise the (marginal) posterior pdf and profile likelihood so that ${\rm max} (\mathcal{P}_{\rm mar})={\rm max} (\mathcal{L}_{\rm prof})=1$.}

\subsection{Priors}\label{sec:priors}

The main parameters of interest in this work are $m_{\rm
  DM}$ and $\sigma_n^{\rm SI}$.  These are accompanied by a set of
astrophysical and experiment-specific systematic nuisance parameters,
as discussed in section~\ref{sec:likeli}.

When it comes to specifying prior pdfs for $m_{\rm DM}$ and
$\sigma_n^{\rm SI}$, we have very little guidance from theory without
resorting to specific dark matter models.  As long as the dark matter
is cold, massive and weakly enough interacting, pretty much all
combinations of values are {\em a priori} allowed.  It thus appears
reasonable to impose uniform priors on both $\log m_{\rm DM}$ and 
$\log \sigma_n^{\rm SI}$. With the
assumption that the dark matter particle is a WIMP, we can at least
roughly confine our prior region. For definiteness, 
we take $\log( m_{\rm DM}/{\rm GeV})$ to lie in the range $0 \to 3$ and allow $\log(\sigma_n^{\rm
  SI}/{\rm cm}^2)$ to vary between $-46 \to -38$, as
reported in table~\ref{tab:prior1}.

Interestingly, the choice of prior boundaries on $m_{\rm DM}$ and
$\sigma_n^{\rm SI}$ also translates directly to how likely we deem the
direct detection experiments to actually make a positive detection.
Consider for instance the loss of detection sensitivity for large DM
masses (due to the large mass splitting between the DM particle and
the nucleus), or for very light WIMPs (because of the energy
threshold):  the larger the prior-space in the \{$m_{\rm
  DM},\sigma_n^{\rm SI}$\}-plane, the smaller the relative fraction that
the experiments will be able to constrain, and the smaller the
subjective prior probability for them to see something.

Our priors for the astrophysical parameters $M_{\rm vir}$ and $c_{\rm
  vir}$ are listed in table~\ref{tab:prior2}.  The ranges for the
concentration parameters are inferred from
simulations~\cite{Navarro:2008kc} for the NFW and Einasto profiles,
and from fits the rotation curves of galaxies for Cored isothermal and Burkert
profiles~\cite{Jimenez:2002vy}. 
Note that since $M_{\rm vir}$ is well-constrained by measurements (see
table~\ref{tab:prior3}), the likelihood at the prior boundaries is negligible, and the inferred 
posterior pdf will be independent of our exact  choice of prior boundaries for this parameter.

\subsection{Numerical implementation and construction of parameter constraints}\label{sec:cl}

We employ a modified version of  the public MCMC  code \texttt{CosmoMC}~\cite{Lewis:2002ah,cosmomc_notes}, 
which uses the Metropolis--Hastings algorithm~\cite{Metropolis53,Hastings70}
to sample the posterior over the full parameter space. 
 The resulting chains are analysed with an adapted version of the accompanying package
  \texttt{GetDist}, supplemented with \texttt{matlab} scripts from the package \texttt{SuperBayeS}~\cite{superbayes,Trotta:2008bp}. 
  One- or two-dimensional marginal posterior pdfs are obtained
from the chains by dividing the relevant parameter subspace into bins and
counting the number of samples per bin.  An $x\%$ credible interval or region containing $x\%$ of the total volume of
$\mathcal{P}_{\rm mar}$ is then constructed by demanding that  $\mathcal{P}_{\rm mar}$  at any point inside the region 
be larger than at any point outside.  In the one-dimensional case, a credible interval thus constructed 
corresponds to the Minimal Credible Interval of~\cite{Hamann:2007pi}. Our profile likelihoods are also computed 
using \texttt{CosmoMC}, but with a 100-fold increase in the number of likelihood evaluations, 
so as to ensure that the tails of the distributions are well sampled and the true global maximum located.

Provided the data are sufficiently constraining---that is, if the prior pdf is nearly
constant and, under a parameter transformation  $f: \theta \to \theta'$, the mapping $f$ is almost linear
over the parameter region where the likelihood is
large---the marginal posterior typically exhibits very little dependence on
the choice of prior.   For data that can only provide an upper or a lower bound on
a parameter (or no bound at all) however, the properties of the
inferred posterior and the boundaries of credible regions can vary
significantly with the choice of prior as well as its limits $\theta_{\rm min}$
and $\theta_{\rm max}$, making an objective interpretation of the
results rather difficult.  As we shall see in the next section, this
is in fact the case for the inference of credible regions in the \{$m_{\rm DM},\sigma_n^{\rm SI}$\}-plane
from Xenon100, CDMSSi and CDMSGe.

In these cases,  in addition to computing credible intervals from 
the fractional volume of the marginal posterior in the  \{$m_{\rm DM},\sigma_n^{\rm SI}$\}-subspace $\mathcal{P}_{\rm mar}(m_{\rm DM},\sigma_n^{\rm SI}|X)$, 
we also construct intervals based on the volume of the marginal posterior in $S$-space $\mathcal{P}_{\rm mar}(S|X)$,
where $S$ is the expected WIMP signal, using a uniform prior on $S$ with a lower boundary at 
zero~\cite{Helene:1982pb}.
An $x$\% upper bound thus constructed has a well-defined Bayesian interpretation that 
the probability of $S\leq S_x$ is $x$\%.  The limit $S_x$ is then mapped onto the  \{$m_{\rm DM},\sigma_n^{\rm SI}$\}-plane
by identifying those combinations of  $m_{\rm DM}$ and $\sigma_n^{\rm SI}$ with $\mathcal{P}_{\rm mar}(m_{\rm DM},\sigma_n^{\rm SI}|X) =\mathcal{P}_{\rm mar}(S_x|X)$.
An $x$\% contour computed in this manner has the property of being independent of our choice of prior boundaries for $m_{\rm DM}$  and $\sigma_n^{\rm SI}$.  Its drawback, however, is that it has no well-defined  probabilistic interpretation in \{$m_{\rm DM},\sigma_n^{\rm SI}$\}-space.%
\footnote{Clearly, the definition of the $S$-based bound and its associated probabilistic interpretation are contingent to our choice of a uniform prior on $S$;
Had we chosen a different prior a different set of limits would have resulted.  Our motivation for using a uniform prior 
stems from the observation that, for Poisson statistics, a Bayesian limit on $S$ constructed
 in the manner described turns out to have a well-defined interpretation in classical statistics, albeit a coincidental one~\cite{Zech:1988un}.  This means
the $S$-based bounds in this work can also be viewed as examples of the hybrid Bayesian/classical approach discussed in~\cite{Cousins:1991qz}.}
To distinguish these $S$-based credible intervals from the conventional ones based on the volume of $\mathcal{P}_{\rm mar}(m_{\rm DM},\sigma_n^{\rm SI}|X)$, 
we label them with a subscript ``$S$'', e.g., $90_S$\%.


\section{Results}\label{sec:results}

We present our inference results in three parts. In section~\ref{sec:SMcase} we discuss the preferred parameter regions in $m_{\rm DM}$ and $\sigma_{n}^{\rm SI}$ for each experiment assuming the SMH (i.e., fixed astrophysics), after marginalising over the nuisance parameters of the experiments.
In section~\ref{sec:velvar} we vary in addition the WIMP velocity distribution in accordance with the DM density profile defined in section~\ref{sec:vel}, and consider the effect
of uncertainties in the astrophysics parameters on the inferred WIMP parameter values.
Finally in section~\ref{sec:combined} we entertain the possibility of a combined analysis of the DAMA and the CoGeNT data. 

\begin{figure}[t]
\begin{minipage}[t]{0.5\textwidth}
\centering
\includegraphics[width=1.1\columnwidth]{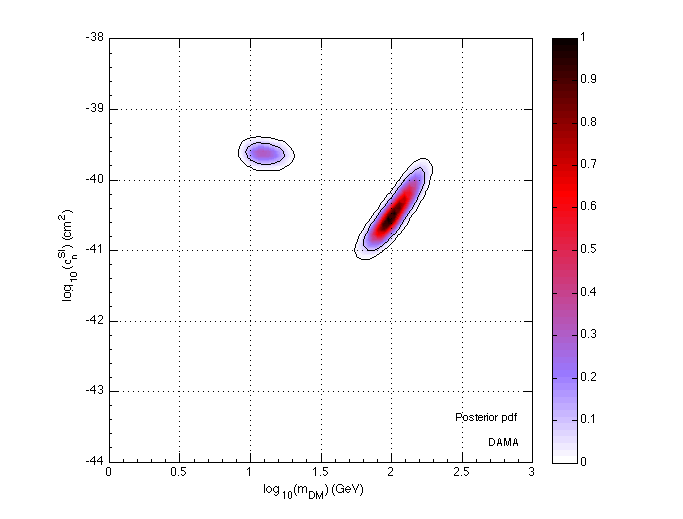}
\end{minipage}
\hspace*{-0.2cm}
\begin{minipage}[t]{0.5\textwidth}
\centering
\includegraphics[width=1.1\columnwidth]{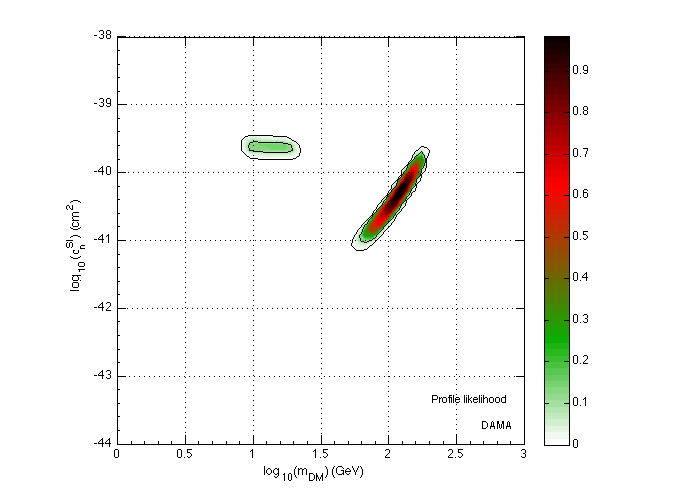}
\end{minipage}
\\
\begin{minipage}[t]{0.5\textwidth}
\centering
\includegraphics[width=1.1\columnwidth]{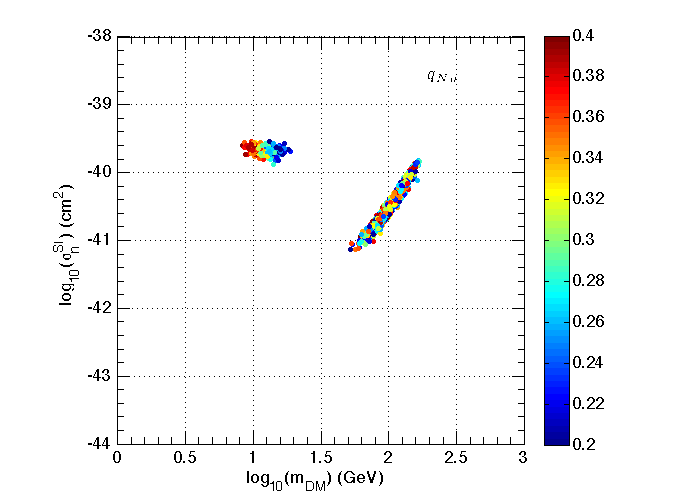}
\end{minipage}
\hspace*{-0.2cm}
\begin{minipage}[t]{0.5\textwidth}
\centering
\includegraphics[width=1.1\columnwidth]{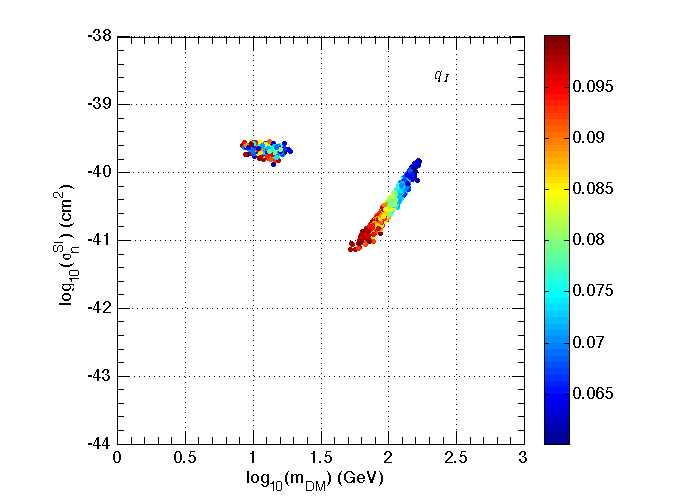}
\end{minipage}
\caption{Inference for DAMA assuming the SMH.
 {\it Top left}: 2D marginal posterior pdf in the \{$m_{\rm DM},\sigma_n^{\rm SI}$\}-plane.  The black solid lines enclose the 90\% and the 99\% credible regions.  {\it Top right}:  Profile likelihood in the \{$m_{\rm DM},\sigma_n^{\rm SI}$\}-plane.   The black solid contours correspond to $\Delta \chi^2_{\rm eff} = 4.6,9.2$.
{\it Bottom left}:  3D marginal posterior pdf for \{$m_{\rm DM},\sigma_n^{\rm SI},q_{\rm Na}$\}, where the $q_{\rm Na}$ direction is represented by the colour code.
{\it Bottom right}: Same as bottom left, but for \{$m_{\rm DM},\sigma_n^{\rm SI},q_{\rm I}$\}.
\label{fig:DamaSMH}}
\end{figure}

\begin{figure}[t]
\hspace*{1.5cm}
\includegraphics[width=0.9\columnwidth]{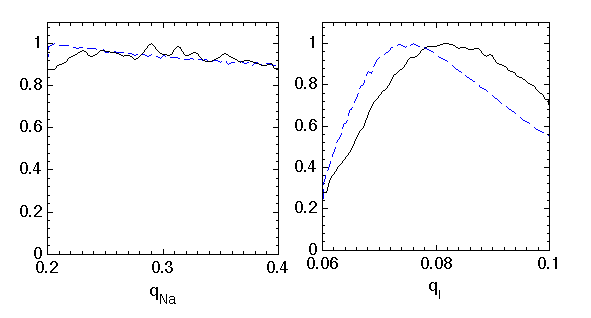}
\caption{
Nuisance parameters for DAMA assuming the SMH. {\it Left}: 1D marginal posterior pdf (black solid line) and profile likelihood (blue dashed line) 
for the quenching factor $q_{\rm Na}$.  {\it Right}: Same as left panel, but for $q_{\rm I}$.
\label{fig:DamaSMHq}}
\end{figure}

\subsection{Standard model halo}\label{sec:SMcase}

\paragraph{DAMA}

Figure~\ref{fig:DamaSMH} shows our inference for the DAMA 36-bin data.  The top panel shows the 2D marginal posterior pdf and the profile likelihood 
in the \{$m_{\rm DM}, \sigma_n^{\rm SI}$\}-subspace, where the two quenching factors $q_{\rm Na}$ and $q_{\rm I}$ have been integrated and profiled out respectively.  Both
approaches single out two preferred islands of parameter space in  \{$m_{\rm DM}, \sigma_n^{\rm SI}$\}.  Moreover, the 
colour coding indicates that $\mathcal{P}_{\rm mar}(m_{\rm DM},\sigma_n^{\rm SI})$ and $\mathcal{L}_{\rm prof}(m_{\rm DM},\sigma_n^{\rm SI})$
coincide to an excellent degree, suggesting that the nuisance directions contribute no strong volume effects.   For the profile likelihood, we also plot two $\Delta \chi^2_{\rm eff}$
contours, defined via
\begin{equation}
\Delta \chi^2_{\rm eff}(m_{\rm DM},\sigma_n^{\rm SI}) \equiv -2 \ln  \mathcal{L}_{\rm prof}(m_{\rm DM},\sigma_n^{\rm SI}) \, ,
\end{equation}
where the choice of  $\Delta \chi^2_{\rm eff}=4.6, 9.2$ coincides with the classical 90\% and 99\% confidence intervals for two degrees of freedom 
(assuming Wilks' theorem holds).   Again, we find remarkable agreement between these contours and  
 the 90\% and 99\% credible regions inferred from the volume of the 2D marginal posterior.  This agreement indicates that when the data are sufficiently informative  so that the likelihood function overcomes the dependence on the priors,
Bayesian and classical statistical methods yield very similar inference results.

The bottom panel of figure~\ref{fig:DamaSMH} illustrates the correlation between  \{$m_{\rm DM}, \sigma_n^{\rm SI}$\} and the quenching factors $q_{\rm Na}$ and $q_{\rm I}$.
As expected, the high mass ($m_{\rm DM} \sim {\cal O}(100)$~GeV) island is insensitive to $q_{\rm Na}$, as indicated by the equal representation of $q_{\rm Na}$ values in the island.  Conversely, 
the low mass ($m_{\rm DM} \sim {\cal O}(10)$~GeV) island shows a strong correlation between $q_{\rm Na}$ and $m_{\rm DM}$, with higher values of $q_{\rm Na}$ favouring the lower masses.
The quenching factor for iodine shows the opposite trend: the low mass island is insensitive to $q_{\rm I}$, while the high mass island finds combinations of
 low $m_{\rm DM}$ and $\sigma_n^{\rm SI}$ values  favoured by large values of  $q_{\rm I}$.  Ultimately, however, 
the DAMA 36-bin data do not constrain either $q_{\rm Na}$ or $q_{\rm I}$, as is evidenced by the fact that all values of $q_{\rm Na}$ and $q_{\rm I}$ allowed by their respective priors are represented in figure~\ref{fig:DamaSMH}.   The same conclusions can be drawn also from figure~\ref{fig:DamaSMHq}, which shows an essentially flat 1D marginal posterior pdf (black solid line) and profile likelihood (dashed blue line) for $q_{\rm Na}$, while for $q_{\rm I}$ one might claim a small preference for $q_{\rm I}=0.07 \to 0.08$ although it is statistically insignificant.

\begin{figure}[t]
\begin{minipage}[t]{0.5\textwidth}
\centering
\includegraphics[width=1.1\columnwidth]{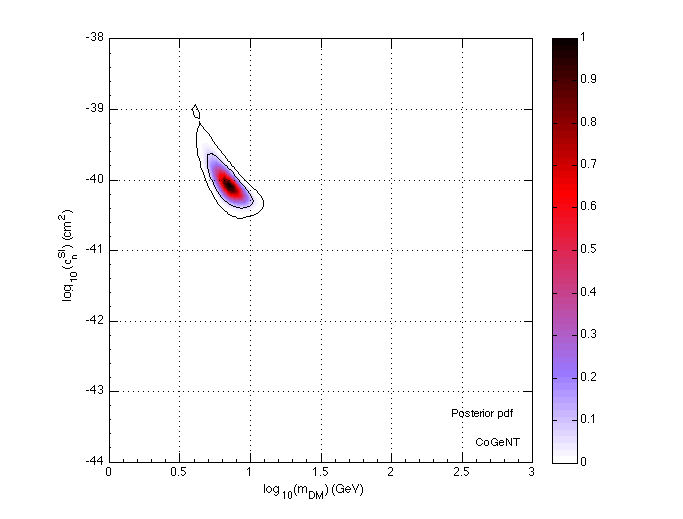}
\end{minipage}
\hspace*{-0.2cm}
\begin{minipage}[t]{0.5\textwidth}
\centering
\includegraphics[width=1.1\columnwidth]{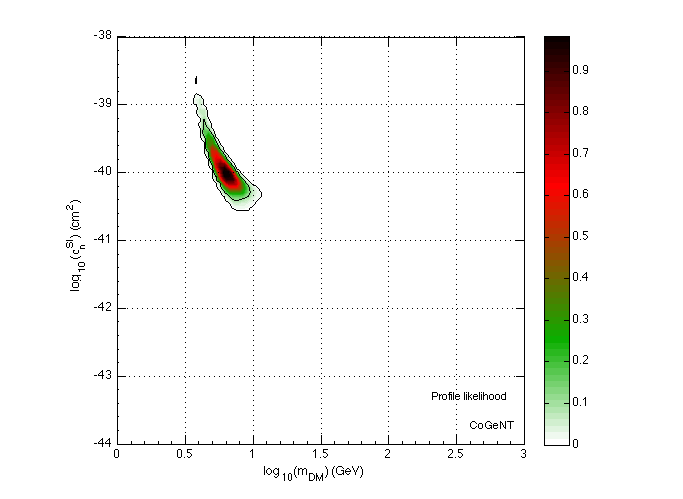}
\end{minipage}
\caption{Inference for CoGeNT assuming the SMH. 
{\it Left}: 2D marginal posterior pdf in the \{$m_{\rm DM},\sigma_n^{\rm SI}$\}-plane.  The black solid lines enclose the 90\% and the 99\% credible regions.
{\it Right}: Profile likelihood in the \{$m_{\rm DM},\sigma_n^{\rm SI}$\}-plane.  The black solid contours correspond to $\Delta \chi^2_{\rm eff} = 4.6,9.2$.
\label{fig:CoGeNTSMH}}
\end{figure}

\paragraph{CoGeNT}
Figure~\ref{fig:CoGeNTSMH} shows the preferred \{$m_{\rm DM}, \sigma_n^{\rm SI}$\}, both in terms of the 2D marginal posterior pdf and the profile likelihood. As in the case of DAMA, the nuisance directions do not contribute strong volume effects, so that 
both the 90\% and 99\% credible regions inferred from the marginal posterior coincide well with the $\Delta \chi^2_{\rm eff}=4.6, 9.2$ contours on the profile likelihood surface,
 and  single out a peak at $m_{\rm DM} \sim 8$~GeV and $\sigma_n^{\rm SI} \sim 10^{-40}~{\rm cm}^2$ as the favoured region.  
The preferred values for the nuisance parameters are reported in table~\ref{tab:cogent}. Our analysis is compatible with all previous analyses of the CoGeNT data, and also with the newest data release~\cite{Frandsen:2011ts}, which claims detection of  an annual modulation  and where the total rate excess leads to a slightly smaller region in the  \{$m_{\rm DM},\sigma_n^{\rm SI}$\}-plane.

\begin{table}[t!]
\caption{1D marginal posterior pdf modes and $90\%$ credible intervals for the CoGeNT nuisance parameters.\label{tab:cogent} }
\begin{center}
\lineup
\begin{tabular}{ll }
\br
Parameter & Preferred value \\
\br
${\cal E}_0$  & $6.1^{+13.1}_{-5.0}$~keV \\
$C$ & $4.0^{+4.8}_{-2.3}$~cpd/kg/keV \\
$G_n$ &  $2.1 \pm 0.5$~cpd/kg/keV \\
\br
\end{tabular}
\end{center}
\end{table}

\begin{figure}[t]
\begin{minipage}[t]{0.5\textwidth}
\centering
\includegraphics[width=1.1\columnwidth]{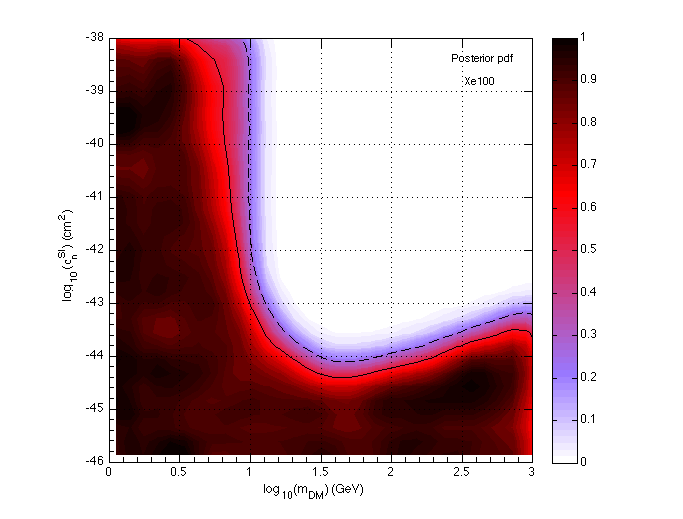}
\end{minipage}
\hspace*{-0.2cm}
\begin{minipage}[t]{0.5\textwidth}
\centering
\includegraphics[width=1.1\columnwidth]{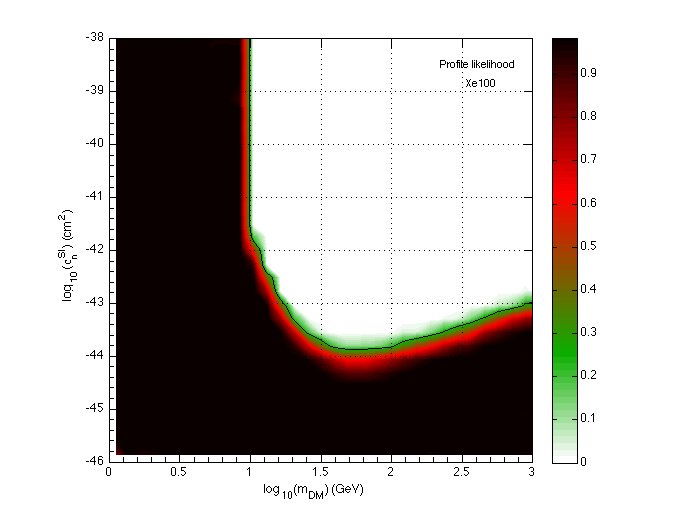}
\end{minipage}
\caption{
Inference for Xenon100 assuming the SMH.
{\it Left}: 2D marginal posterior pdf in the \{$m_{\rm DM},\sigma_n^{\rm SI}$\}-plane.  The black solid line indicates the 90\% bound inferred from the volume of the marginal 
posterior, while the black dashed line denotes the invariant $90_S$\% contour.
{\it Right}: Profile likelihood in the \{$m_{\rm DM},\sigma_n^{\rm SI}$\}-plane.  The black dashed line corresponds to $\Delta \chi^2_{\rm eff} = 2.7$.
\label{fig:Xe100SMH}}
\end{figure}

\begin{figure}[t]
\centering
\includegraphics[width=0.7\columnwidth]{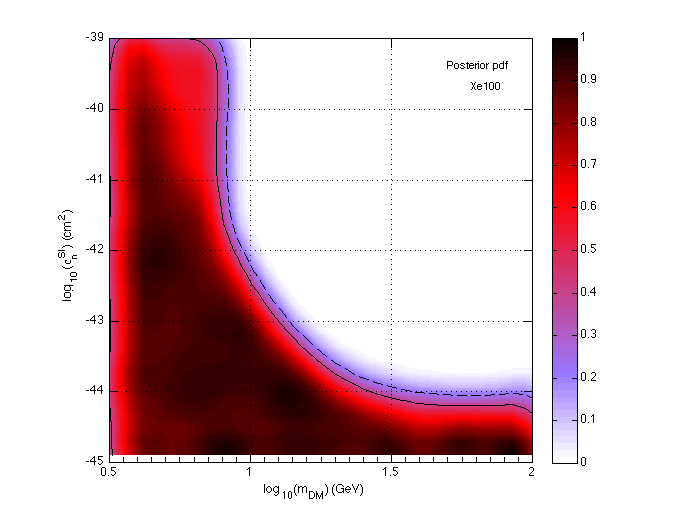}
\caption{2D marginal posterior pdf in the \{$m_{\rm DM},\sigma_n^{\rm SI}$\}-plane 
for Xenon100 assuming the SMH and an alternative set of prior boundaries for $m_{\rm DM}$ and $\sigma_n^{\rm SI}$.  The black solid line corresponds to the 
90\% bound inferred from the volume of the marginal posterior, while the black dashed line represents the invariant $90_S$\% contour.
\label{fig:Xe100mod}}
\end{figure}

\paragraph{Xenon100} 
Our inference results for Xenon100 are shown in figure~\ref{fig:Xe100SMH}. Firstly, we note that both the 2D marginal posterior pdf and the profile likelihood
form a plateau  as $m_{\rm DM}$ and $\sigma_n^{\rm SI}$ approach their respective lower boundaries.%
\footnote{Strictly speaking, the profile likelihood shown in figure~\ref{fig:Xe100SMH} for Xenon100 is a quasi-profile likelihood, computed 
 after the full likelihood function~(\ref{eq:lkhxetot}) has been analytically  marginalised over the background uncertainties.}
In this case, credible regions constructed from the
volume of the marginal posterior in  \{$m_{\rm DM}, \sigma_n^{\rm SI}$\}-space can be strongly dependent on our choice of the $m_{\rm DM}$ and $\sigma_n^{\rm SI}$
prior boundaries.  This is illustrated in the left panel of figure~\ref{fig:Xe100SMH} and in figure~\ref{fig:Xe100mod}.  In both figures the 90\% credible region is demarcated by
the black solid line, except that in figure~\ref{fig:Xe100mod} we have chosen a set of prior boundaries for  $m_{\rm DM}$ and $\sigma_n^{\rm SI}$ ($0.5 \leq \log(m_{\rm DM}/{\rm GeV}) \leq 2$ and $-45 \leq \log(\sigma_n^{\rm SI}/{\rm cm}^2) \leq -39$) differing from the default choices of table~\ref{tab:prior1}.  The discrepancy between the encompassed parameter space is clear. As an example, while the point  \{$\log(m_{\rm DM}/{\rm GeV})=0.8,  \log(\sigma_n^{\rm SI}/{\rm cm}^2) = -40$\} sits outside the 90\% credible region in figure~\ref{fig:Xe100SMH}, it sits comfortably within in figure~\ref{fig:Xe100mod}.

On the other hand, the $90_S$\% bound (black dashed line in left panel of figure~\ref{fig:Xe100SMH} and in figure~\ref{fig:Xe100mod})
is clearly independent of the boundary conditions as discussed in section~\ref{sec:cl}, and the parameter region enclosed compares well with the 
$\Delta \chi_{\rm eff}^2 \leq 2.7$  (or $S \leq 5.2$) region  in the profile likelihood (right panel of figure~\ref{fig:Xe100SMH}).   We will therefore use the $90_S$\% bound 
 in the following discussion.

Our exclusion limit on $\sigma_n^{\rm SI}$ at high WIMP masses ($m_{\rm DM} \gwig 30$~GeV) agrees very well with that provided by the Xenon100 collaboration~\cite{:2011hi}.
However, at low WIMP masses, our bound on $m_{\rm DM}$  is much less constraining compared with all previous analyses~\cite{:2011hi,Andreas:2010dz,Savage:2010tg,Schwetz:2010gv,Farina:2011bh}.  This is clearly a consequence of  the uncertainties in the scintillation efficiency ${\rm L}_{\rm eff}(E)$ in the low recoil energy 
($1<E/{\rm keVnr}<3$) region, which we have accounted for in this work using the nuisance parameter $m$.%
\footnote{ We note that the exclusion limits reported by the Xenon collaboration in~\cite{:2011hi} and \cite{Aprile:2011hx}  are in fact 1D limits on $\sigma_m^{\rm SI}$ for {\it fixed}  values 
of $m_{\rm DM}$.  These limits are naturally different from our 2D limits for $\{m_{\rm DM},\sigma_n^{\rm SI}\}$, which come from considering the joint probability distribution of $m_{\rm DM}$ and $\sigma_n^{\rm SI}$.}
The preferred value for this parameter is $m=0.07\pm 0.04$ (90\% C.I.), which corresponds to a marginal preference for a gentler slope for  $\rm L_{eff}(E)$ at $1<E/{\rm keVnr}<3$
with respect to the Xenon100 collaboration's best-fit.

\begin{figure}
\centering
\includegraphics[width=0.7\columnwidth]{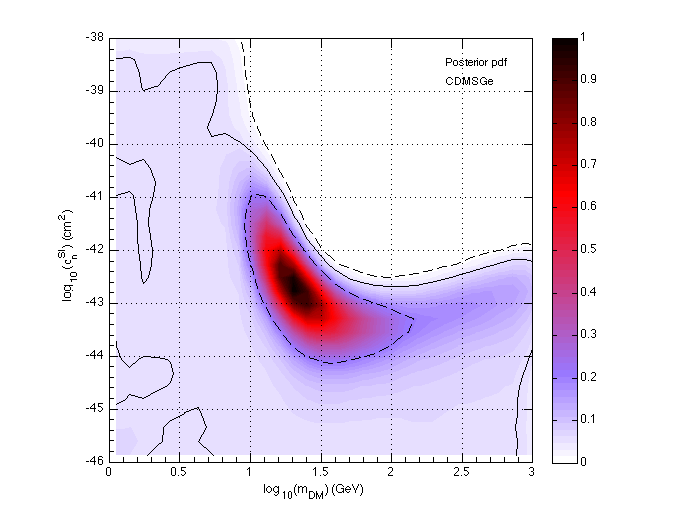}
\caption{Posterior pdf for CDMSGe assuming the SMH in the \{$m_{\rm DM},\sigma_n^{\rm SI}$\}-plane. The black solid line indicates the 90\% bound inferred from the volume of the posterior, while the black dashed lines denote the invariant $90_S$\% and $99_S$\% contours (corresponding to $\Delta \chi^2_{\rm eff}=3,7.4$).
\label{fig:CDMSGeSMH}}
\end{figure}

\paragraph{CDMSGe}
The posterior pdf as a function of $m_{\rm DM}$ and $\sigma_n^{\rm SI}$ is shown in figure~\ref{fig:CDMSGeSMH}.  Since there are no nuisance parameters---besides the background uncertainty which we have already marginalised analytically in order to obtain the effective likelihood~(\ref{eq:cdmsgeeff}), the  posterior pdf in  figure {\it is} the full posterior pdf of the problem.  It also coincides with the effective likelihood~(\ref{eq:cdmsgeeff}) because of our choice of uniform priors.
 A peak can be seen at a DM mass of 23~GeV and a cross-section of $9 \times 10^{-44} \ {\rm cm^2}$.   While this is a tantalising hint, a detection cannot be called because the probability density is still significant at much of 
the prior boundaries ($\mathcal{P} \sim 0.1$).

It then remains for us to set an exclusion limit in the  \{$m_{\rm DM},\sigma_n^{\rm SI}$\}-plane.  The 90\% contour inferred from the volume of posterior (black solid line)
forms a semi-closed region subject strongly to our choice of prior boundaries.   The invariant $90_S$\% or $\Delta \chi_{\rm eff}^2 \leq 3.0$ region (black dashed line), however,  is a closed island  in the  \{$m_{\rm DM},\sigma_n^{\rm SI}$\}-plane, while  the $99_S$\%  ($\Delta \chi_{\rm eff}^2 =7.4$) contour indicates an exclusion limit.

Compared with the analysis in figure~3 of~\cite{Kopp:2009qt}, our posterior pdf/likelihood appears to be more strongly peaked relative to the plateau, leading 
to a closed $90_S$\% region while~\cite{Kopp:2009qt} finds an open one.  At the same time, our peak region appears to be much broader than that of~\cite{Kopp:2009qt},
so that their 90\% contour runs right into our peak region where the posterior pdf/likelihood is still high ($>0.5$). Fixing the number of background events to the mean value, i.e., setting $\sigma_B=0$ in the effective likelihood~(\ref{eq:cdmsgeeff}), does not ameliorate the discrepancy.
Since reference~\cite{Kopp:2009qt} does not specify the likelihood function used in their analysis, we have no more handle to trace the origin  of the disagreement.

\begin{figure}
\centering
\includegraphics[width=0.7\columnwidth]{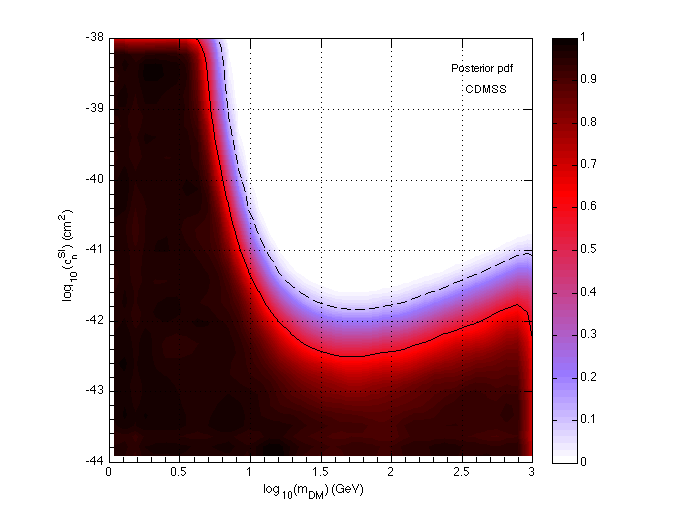}
\caption{Posterior pdf for CDMSSi assuming the SMH
 in the \{$m_{\rm DM},\sigma_n^{\rm SI}$\}-plane. The black solid line indicates the 90\% bound inferred from the volume of the posterior, while the black dashed line denotes the invariant $90_S$\% contour (corresponding to $\Delta\chi^2_{\rm eff}=4.2$).
\label{fig:CDMSSiSMH}}
\end{figure}

\paragraph{CDMSSi}
The analysis of the CDMSSi data is summarised in figure~\ref{fig:CDMSSiSMH}.   Similar to CDMSGe, the posterior pdf presented in the figure is the full posterior pdf of the problem (barring analytic marginalisation over the background) and coincides with the effective likelihood~(\ref{eq:cdmssieff}).
As in the case of Xenon100,  the CDMSSi data are not sufficiently constraining to isolate a preferred region, so that the 90\% credible region inferred from the posterior volume (black solid line) depends on our choice of prior boundaries.  On the other hand, the $90_S\%$ region (black dashed line, corresponding to $\Delta\chi^2_{\rm eff} \leq4.2$ or $S \leq 3.3$) is independent of the $m_{\rm DM}$ and $\sigma_n^{\rm SI}$ prior boundaries,  
and agrees well with the exclusion limit constructed by the CDMS collaboration~\cite{Akerib:2005kh}.

\begin{figure}
\centering
\includegraphics[width=0.9\columnwidth]{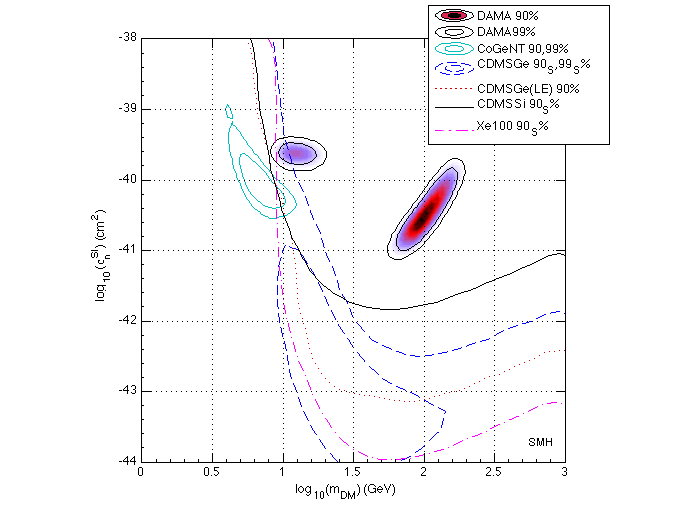}
\caption{2D credible regions for the individual experimental bounds and regions assuming the SMH, combined in a single plot. For DAMA (shaded) and CoGeNT (cyan) we show the 90\% and 99\% contours.
The black solid line represents the $90_S$\% bound for CDMSSi, and the pink dot-dash curve for Xenon100.  For CDMSGe we show both the $90_S$\% and 
$99_S$\% contours in blue dashed lines, while the red dotted line is the 90\% contour for CDMSGe(LE) corresponding to $\Delta\chi^2_{\rm eff} = 4.6$.
\label{fig:allSMH}}
\end{figure}

\paragraph{SMH state of the art}
We summarise our results for fixed astrophysics in figure~\ref{fig:allSMH}, in which we show all experimental constraints in one plot.  
For DAMA and CoGeNT we indicate the 90\% and 99\% credible regions, while for the exclusion limits of the other three experiments we show the invariant  $90_S\%$ contours (also $99_S\%$ for CDMSGe).

We find that the parameter region favoured by DAMA is incompatible with the  $90_S\%$ credible regions of Xenon100 and CDMSSi, and partially allowed by the $99_S\%$ region of  CDMSGe. In contrast,  the CoGeNT preferred region is only marginally incompatible with these exclusion limits.  Of particular interest is
the compatibility between CoGeNT and Xenon100.  While the Xenon100 collaboration claims that their exclusion limit has ruled out the CoGeNT preferred region~\cite{:2011hi}, we find that when uncertainties in the scintillation efficiency ${\rm L}_{\rm eff}(E)$ at low recoil energies are accounted for, the CoGeNT and the Xenon100 data can 
find some common ground.

Between CoGeNT and DAMA we find that their 99\% credible regions do not overlap, despite marginalisation over the quenching factors $q_{\rm Na}$ and $q_{\rm I}$ 
for DAMA.  This is a consequence of our choice of prior boundaries for $q_{\rm Na}$ ($0.2 \to 0.4$),   especially in view of~\cite{Schwetz:2010gv,Hooper:2010uy}, where it has been suggested that in order to make  DAMA and CoGeNT compatible large quenching factors for sodium (e.g., $q_{\rm Na}=0.6$) and for germanium should be considered.  
Allowing up to 10\% of channelling for DAMA could also improve the agreement between the two experiments, by shifting the DAMA low mass region downwards in the \{$m_{\rm DM},\sigma_n^{\rm SI}$\}-plane~\cite{Bozorgnia:2010xy}.

Lastly, we also show the exclusion bound derived from  CDMSGe(LE) (red dotted line in figure~\ref{fig:allSMH}).
 Since the likelihood function~(\ref{eq:cdmsgele}) for this case 
is a multivariate gaussian,  we can infer an invariant 90\% exclusion bound similar to the $90_S\%$ bound
by demanding that $\Delta\chi_{\rm eff}^2<4.6$.  At low masses this 90\% bound turns out to be very close to the CDMSSi exclusion limit, so that the DAMA preferred region falls 
outside the credible region, while the CoGeNT region falls mostly within.
The main difference between this low energy analysis and the standard CDMSGe is that the former does not find any closed region at small WIMP masses. 
Compared with other experiments, for masses larger than 10~GeV the Xenon100 bound is more constraining.  We therefore do not consider the CDMSGe(LE) exclusion limit any further.

\subsection{Variable WIMP velocity distribution and astrophysics}\label{sec:velvar}

\begin{figure}
\begin{minipage}[t]{0.5\textwidth}
\centering
\includegraphics[width=1.1\columnwidth]{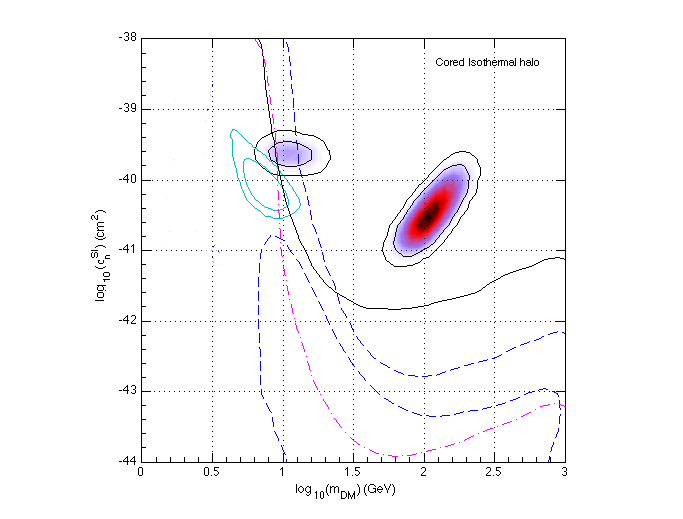}
\end{minipage}
\hspace*{-0.2cm}
\begin{minipage}[t]{0.5\textwidth}
\centering
\includegraphics[width=1.1\columnwidth]{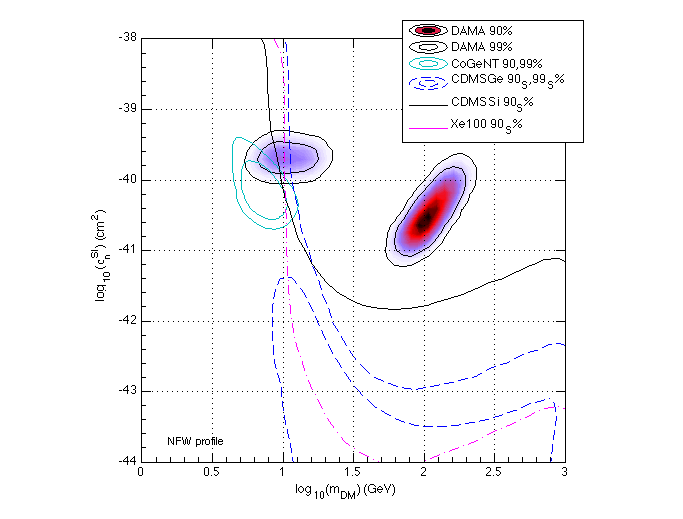}
\end{minipage}
\\
\begin{minipage}[t]{0.5\textwidth}
\centering
\includegraphics[width=1.1\columnwidth]{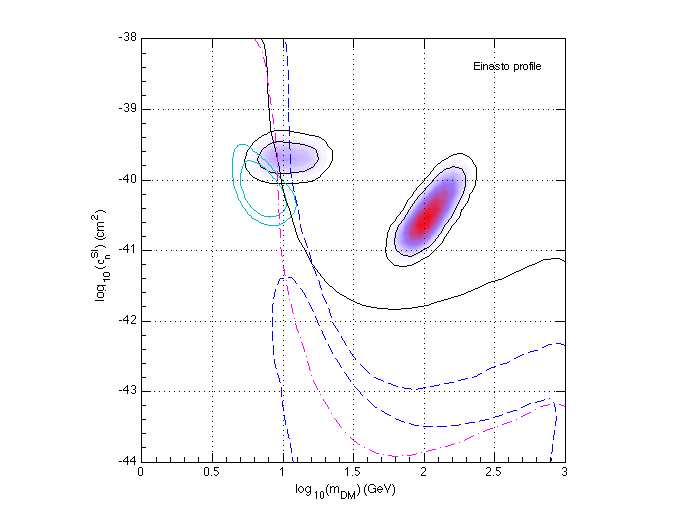}
\end{minipage}
\hspace*{-0.2cm}
\begin{minipage}[t]{0.5\textwidth}
\centering
\includegraphics[width=1.1\columnwidth]{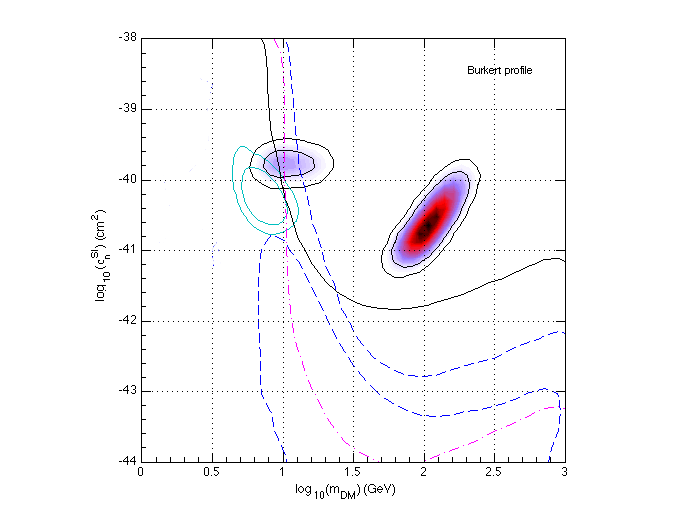}
\end{minipage}
\caption{Same as figure~\ref{fig:allSMH}, but with variable astrophysics, assuming the Cored isothermal (top left), 
NFW (top right), Einasto (bottom left), and Burkert (bottom right) profiles.
\label{fig:all}}
\end{figure}

Figure~\ref{fig:all} shows the effects of astrophysical uncertainties on the inferred \{$m_{\rm DM},\sigma_n^{\rm SI}$\} parameter space, for each of the four DM density profiles
considered (see section~\ref{sec:haloprofiles}).
 The corresponding 
preferred values for the local dark matter density, the circular and the escape velocities are reported in table~\ref{tab:allparams}.

\begin{table}[t!]
\small
\caption{1D posterior pdf modes and $90\%$ credible intervals for the circular velocity $v_0$,
escape velocity $v_{\rm esc}$, and the local DM density $\rho_{\odot}$ for DM density profiles considered in this work. \label{tab:allparams}}
\begin{center}
\lineup
\begin{tabular}{ l | lll }
\br
& $v_0$ (${\rm  km \ s}^{-1}$) & $v_{\rm esc}$ (${\rm km \ s}^{-1}$) & $\rho_{\odot}$  ($ {\rm GeV \ cm}^{-3}$) \\ 
\br
\bfseries{Cored Isothermal} & & &\\
DAMA & $210^{+ 26}_{-16}$ & $628_{-17}^{+22}$ &$0.31_{-0.03}^{+0.05}$ \\
CoGeNT & $209^{+14}_{-21}$  & $628 \pm 18$  & $0.31 \pm 0.04$ \\
CDMSGe & $208_{-16}^{+22}$ & $628_{-21}^{+23}$ & $0.31 \pm 0.05$  \\
CDMSSi & $210_{-16}^{+29} $ &$ 628 \pm 21$ & $ 0.31^{+0.05}_{-0.04}$ \\
Xenon100 &$211_{-19}^{+26}$ &$629 \pm 21$ &$0.31\pm 0.04$  \\
\mr
{\bfseries NFW} & & & \\
DAMA & $220^{+40}_{-20}$ & $558_{-16}^{+19}$ & $0.37_{-0.09}^{+0.15}$ \\
CoGeNT & $219_{-18}^{+38}$ & $559 \pm 17$ & $0.37_{-0.08}^{+0.20}$ \\
CDMSGe & $218_{-18}^{+41}$ & $559 \pm 18$ & $0.37_{-0.08}^{+0.16}$ \\
CDMSSi & $218_{-19}^{+44}$ & $560_{-18}^{+19}$ & $0.36_{-0.09}^{+0.18}$ \\
Xenon100 &$219_{-20}^{+43}$ &$559\pm 18$ & $0.37_{-0.08}^{+0.16}$ \\
\mr
{\bfseries Einasto} & & & \\
DAMA & $221_{-19}^{+39}$ & $560_{-18}^{+13}$ & $0.36_{-0.08}^{+0.14}$ \\
CoGeNT & $222_{-19}^{+42}$ & $562_{-21}^{+11}$ & $0.36_{-0.08}^{+0.15}$ \\
CDMSGe &$221_{-19}^{+44}$ & $561_{-22}^{+11}$ & $0.36_{-0.08}^{+0.15}$ \\
CDMSSi & $221_{-19}^{+44}$ & $ 561_{-22}^{+11}$ & $0.36_{-0.08}^{+0.15}$ \\
Xenon100 & $221_{-19}^{+44}$ & $562_{-22}^{+11}$ & $0.36_{-0.08}^{+0.15}$ \\
\mr
{\bfseries Burkert} & & &\\
DAMA & $214_{-21}^{+36}$ & $548_{-16}^{+29}$ & $0.44_{-0.12}^{+0.16}$ \\
CoGeNT & $216_{-22}^{+35}$ & $550 \pm 20$ & $0.44_{-0.12}^{+0.16}$ \\
CDMSGe & $215_{-23}^{+35}$ & $549 \pm 19$ & $0.44_{-0.12}^{+0.18}$ \\
CDMSSi & $215_{-23}^{+35}$ & $550 \pm 22$ & $0.44_{-0.13}^{+0.18}$ \\
Xenon100 & $216_{-23}^{+35}$ & $550 \pm 21$ & $0.44_{-0.13}^{+0.16} $ \\
\br
\end{tabular}
\end{center}
\end{table}

Firstly, we note that all four DM profiles give very similar inference results on the \{$m_{\rm DM},\sigma_n^{\rm SI}$\}-plane.  This means  that the exact shape of the DM halo density profile---at least within the class of spherically symmetric, smooth profiles---does not yet play a role in direct DM searches.
This conclusion is further supported by the inferred local DM density, circular and escape velocities presented in table~\ref{tab:allparams}.  The preferred values for these quantities  differ from profile to profile, with the Cored isothermal halo in particular favouring the very high end of the observationally allowed 
escape velocities (see table~\ref{tab:prior2}). However, once the DM halo profile has been fixed, we see that the preferred values for $v_0$, $v_{\rm esc}$ and $\rho_\odot$ and their associated uncertainties 
are virtually independent of the additional constraints from the DM experiments.  In other words, direct DM searches are not at the moment contributing towards constraining the astrophysics of the problem.

\begin{figure}[t]
\begin{minipage}[t]{0.33\textwidth}
\centering
\includegraphics[width=1.15\columnwidth]{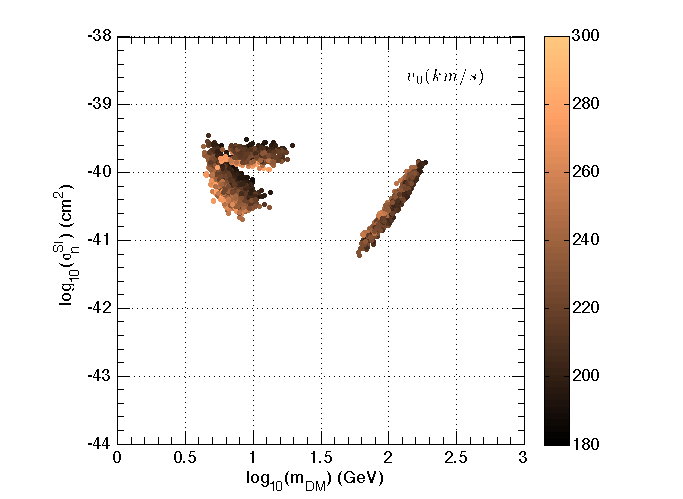}
\end{minipage}
\begin{minipage}[t]{0.33\textwidth}
\centering
\includegraphics[width=1.15\columnwidth]{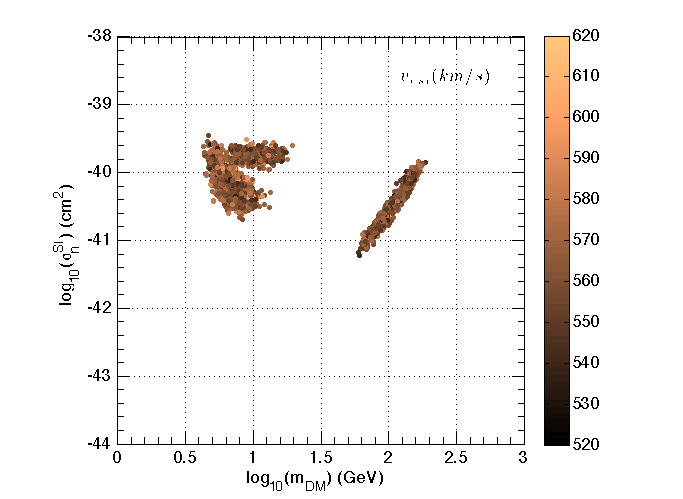}
\end{minipage}
\begin{minipage}[t]{0.33\textwidth}
\centering
\includegraphics[width=1.15\columnwidth]{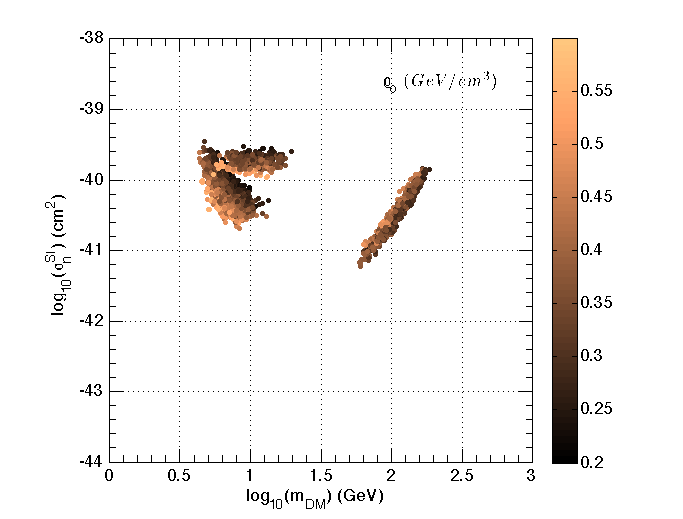}
\end{minipage}
\caption{3D marginal posterior pdf  for DAMA and CoGeNT for \{$m_{\rm DM},\sigma_n^{\rm SI}\}$ and the circular velocity $v_0$ (left), the 
escape velocity $v_{\rm esc}$ (centre), and the local DM density $\rho_\odot$ (right), assuming the NFW profile.  
The third parameter direction is represented by the colour code.
\label{fig:CoDA3D}}
\end{figure}

Secondly,  we note that  allowing for uncertainties in the astrophysics significantly expands the closed regions of DAMA, CDMSGe and CoGeNT, while
the exclusion limits  tend to shift a little to the right. For all four profiles, the preferred regions of DAMA and CoGeNT now appear to marginally overlap: for 
the NFW, Einasto and Burkert profiles we see an overlap between the 90\% credible region of DAMA with the 99\% region of CoGeNT and vice versa, while for 
the Cored isothermal the agreement is a little worse. One may be tempted to claim some degree of agreement between DAMA and CoGeNT based on this partial overlap.  However, before we do so, it is important that we also examine the degree of overlap between the preferred regions in the other parameter directions.

Figure~\ref{fig:CoDA3D} shows the 3D marginal posterior pdf for \{$m_{\rm DM},\sigma_n^{\rm SI}$\} and a third parameter direction  $v_0$, $v_{\rm esc}$ and $\rho_\odot$.
Here,  we see that while it is not impossible to find a value of $v_{\rm esc}$ that satisfies both DAMA and CoGeNT simultaneously, there is a clear trend that  combinations 
of larger \{$m_{\rm DM},\sigma_n^{\rm SI}$\} values tend to prefer higher values of $v_0$ (and similarly for $\rho_\odot$).  This indicates that although DAMA and CoGeNT appear 
to overlap in the  \{$m_{\rm DM},\sigma_n^{\rm SI}$\}-plane, there is in fact very little overlap between them in the $v_0$ direction (and naturally also in the $\rho_\odot$ direction which enters into the differential recoil rates as a common normalisation factor).

\begin{figure}
\begin{minipage}[t]{0.5\textwidth}
\centering
\includegraphics[width=1.1\columnwidth]{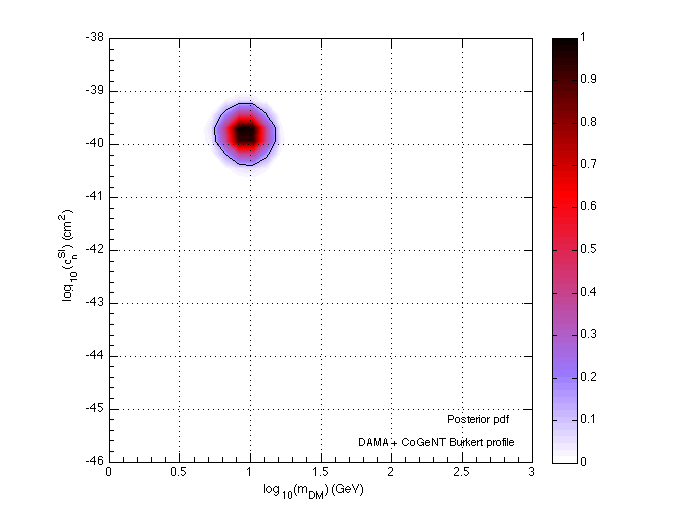}
\end{minipage}
\hspace*{-0.2cm}
\begin{minipage}[t]{0.5\textwidth}
\centering
\includegraphics[width=1.1\columnwidth]{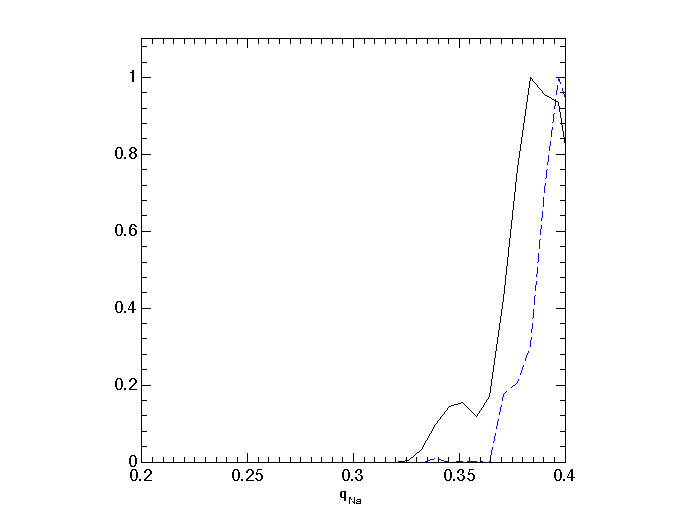}
\end{minipage}
\caption{Inference from the combined DAMA and CoGeNT fit assuming the Burkert halo profile.  {\it Left}: 2D marginal posterior pdf in the  \{$m_{\rm DM},\sigma_n^{\rm SI}$\}-plane. The black solid lines correspond to the 90\% and 99\% bound inferred from the volume of the marginal posterior. {\it Right}: 1D marginal posterior pdf (black solid line) and profile likelihood (blue dashed line) for the quenching factor $q_{\rm Na}$. 
\label{fig:Combined}}
\end{figure}

\subsection{Combined fit?}~\label{sec:combined}

Despite apparent difficulties to reconcile the DAMA and the CoGeNT preferred regions within the boundaries of our nuisance and astrophysics models, let us for a moment entertain the possibility of a combined fit. Figure~\ref{fig:Combined} shows the preferred region in the \{$m_{\rm DM}, \sigma_n^{\rm SI}$\}-plane
from a combined fit of DAMA and CoGeNT assuming the Burkert profile (marginalised over all nuisance and astrophysics parameters as usual).    The corresponding 1D marginal credible intervals for the nuisance and astrophysics parameter are displayed in table~\ref{tab:aastrocomb}.

The best-fit point of the combined fit corresponds to  a mass of $9.2$~GeV and a cross-section of $1.26 \times 10^{-40}\ {\rm cm}^2$.  However, this fit comes at the expense 
of a significant shift in the circular velocity: $v_0=176^{+33}_{-1}  \ {\rm km \ s}^{-1}$ (90\% C.I.)  from the combined fit, versus  $v_0=214^{+36}_{-21}  \ {\rm km \ s}^{-1}$
from fitting either DAMA or CoGeNT alone.  The preferred local DM density $\rho_\odot$ and escape velocity $v_{\rm esc}$ also suffer a downward shift respectively, although not a significant one in either case.  For the nuisance parameters, we find that for CoGeNT, the normalisation for the exponentially decaying background $C$ has come down a little, and the decay rate ${\cal E}_0$ is significantly more constrained.  The radiative peaks, on the other hand, are a well-defined feature, and consequently the preferred value 
for their height $G_n$ has not been affected by the combined fit.

Most interestingly, we find that the DAMA sodium quenching factor $q_{\rm Na}$, which was previously an unconstrained quantity (see figure~\ref{fig:DamaSMHq}),
now shows a preference for high values (right panel of figure~\ref{fig:Combined}).  In particular, the 1D profile likelihood (blue dashed line) has no local maximum, and hits its highest point right at the prior boundary $q_{\rm Na}=0.4$.  This suggests that if we had allowed for a wider prior range for $q_{\rm Na}$, an even higher value might have been preferred.
This result is consistent with previous suggestions that a higher value for $q_{\rm Na}$ could 
improve the compatibility of DAMA and CoGeNT (see section~\ref{sec:qna}).

\begin{figure}
\begin{minipage}[t]{0.5\textwidth}
\centering
\includegraphics[width=1.03\columnwidth]{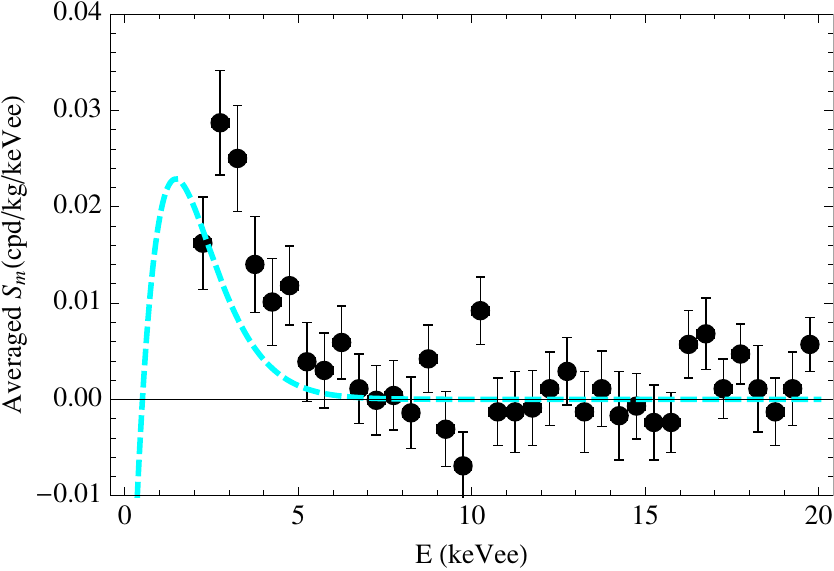}
\end{minipage}
\hspace*{+0.2cm}
\begin{minipage}[t]{0.5\textwidth}
\centering
\includegraphics[width=\columnwidth]{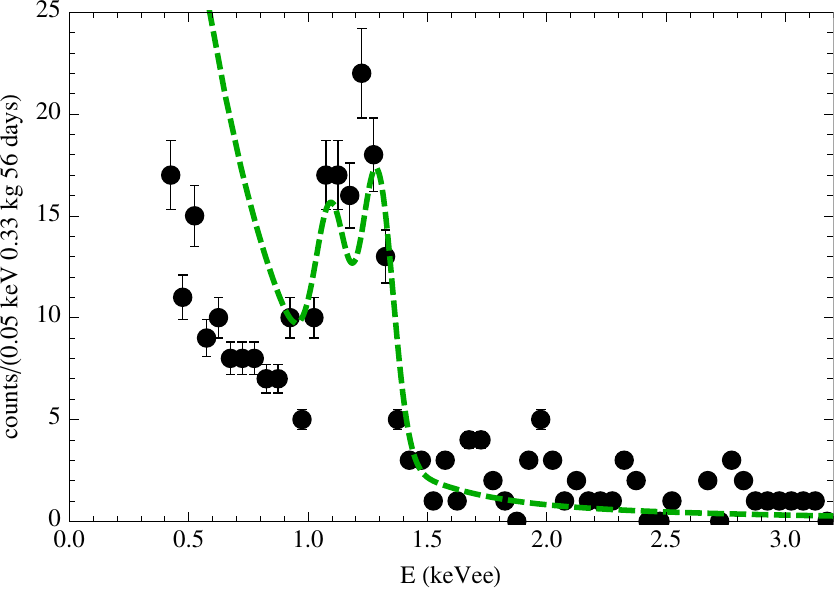}
\end{minipage}
\caption{{\it Left}: The expected signal for DAMA using the best-fit point of our combined DAMA/CoGeNT fit.  {\it Right}: The expected signal for 
CoGeNT using the best-fit point of our combined fit.
\label{fig:fitbyeye}}
\end{figure}

To assess the quality of the fit for the best-fit point singled out by the combined run, we look at the spectral shape of the expected signal in both detectors. In figure~\ref{fig:fitbyeye} we show on the left the averaged modulated amplitude of DAMA, and on the right the number of counts per bin versus the recoil energy for CoGeNT. Superimposed here are the predictions for the best-fit point (dashed lines).  Clearly, the ``best-fit'' point is actually a bad fit for both experiments.  For the best-fit DM mass, cross-section and nuisance parameters, a better fit in DAMA would be obtained by increasing both $v_0$, which would shift the spectral curve to the left, and $\rho_\odot$, which would result in a global enhancement of the signal. For CoGeNT the trend is the opposite: a better fit is obtained by decreasing $\rho_\odot$ and increasing $v_0$, as demonstrated in figure~\ref{fig:CoDA3D}. For both detectors the signal is rather insensitive to the value of $v_{\rm esc}$.

\begin{table}[t!]
\caption{1D marginal posterior pdf modes and $90\%$ credible intervals for the astrophysical 
and the nuisance parameters from the combined DAMA and CoGeNT fit 
assuming the Burkert profile.\label{tab:aastrocomb} }
\begin{center}
\lineup
\begin{tabular}{ll }
\br
Parameter & Preferred value \\
\br
$v_0$  & $176 ^{+33}_{-1} \ {\rm km \ s}^{-1}$ \\
$v_{\rm esc}$ & $533^{+27}_{-8} \  {\rm km \ s}^{-1}$ \\
$\rho_{\odot}$ & $0.3^{+0.2}_{-0.09} \ {\rm GeV \ cm}^{-3}$ \\
$q_{\rm Na} $ &$ 0.38^{+0.02}_{-0.03}$\\
${\cal E}_0$  & $5 \pm 1.2$~keV \\
$C$ & $2.8^{+2.8}_{-1.7}$~cpd/kg/keV \\
$G_n$ & $2.2\pm 0.4$~cpd/kg/keV \\
\br
\end{tabular}
\end{center}
\end{table}

\subsection{A larger $q_{\rm Na}$\label{sec:qna}}

As suggested in the previous section, allowing for a larger sodium quenching factor $q_{\rm Na}$ for DAMA may improve the combined DAMA/CoGeNT fit.  We explore this possibility here by raising the upper limit of our 
prior range on $q_{\rm Na}$ from 0.4 to 0.6, and recomputing the preferred regions for combined DAMA/CoGeNT assuming a Burkert profile.

\begin{figure}
\begin{minipage}[t]{0.5\textwidth}
\centering
\includegraphics[width=1.1\columnwidth]{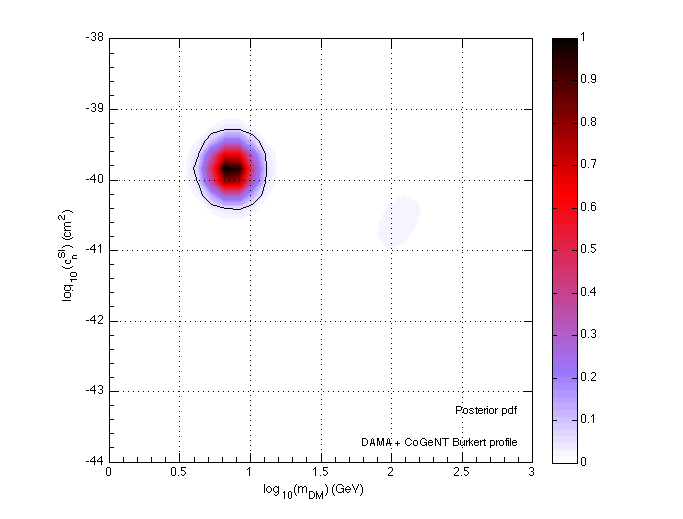}
\end{minipage}
\hspace*{-0.2cm}
\begin{minipage}[t]{0.5\textwidth}
\centering
\includegraphics[width=1.1\columnwidth]{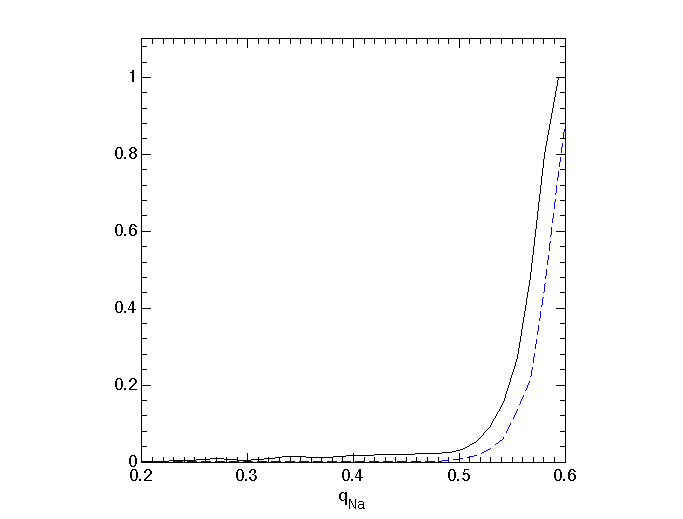}
\end{minipage}
\caption{Same as figure~\ref{fig:Combined}, but for an extended prior range for the DAMA sodium quenching factor $q_{\rm Na}$ (up to $q_{\rm Na} = 0.6$). 
\label{fig:Combined1}}
\end{figure}

The results are shown in figure~\ref{fig:Combined1}. On the left panel, the preferred region in the \{$m_{\rm DM}, \sigma_n^{\rm SI}$\}-plane is similar to that inferred using our standard prior on $q_{\rm Na}$ 
(see figure~\ref{fig:Combined}), with the best-fit point now corresponding to a mass of $7.38$~GeV and a cross-section of $9.64 \times 10^{-41}\ {\rm cm}^2$.  
On the right panel, we see that both the 1D marginal posterior pdf (black solid) and profile likelihood (dashed blue) for $q_{\rm Na}$ rise sharply between $q_{\rm Na}= 0.5$ and $0.6$, hitting their highest points at the edge of the prior boundary. This confirms the trend that the combined DAMA and CoGeNT data prefer a large $q_{\rm Na}$.

Interestingly, the preferred values for the astrophysical parameters from the extended combined fit are now more closely in line with those from fits to the individual experiments alone
(see table~\ref{tab:aastrocomb1}).  The most significant change can be seen for the circular velocity, which now has the preferred value $v_0 = 201^{+35}_{-17}\ {\rm km \ s}^{-1}$ (90\% C.I.),  in contrast to 
(a)  $v_0=176^{+33}_{-1}  \ {\rm km \ s}^{-1}$ from the standard combined fit in section~\ref{sec:combined}, and (b) $v_0=214^{+36}_{-21}  \ {\rm km \ s}^{-1}$ and  $v_0=216^{+35}_{-22}  \ {\rm km \ s}^{-1}$ from
DAMA and CoGeNT alone respectively.

\begin{figure}
\begin{minipage}[t]{0.5\textwidth}
\centering
\includegraphics[width=1.03\columnwidth]{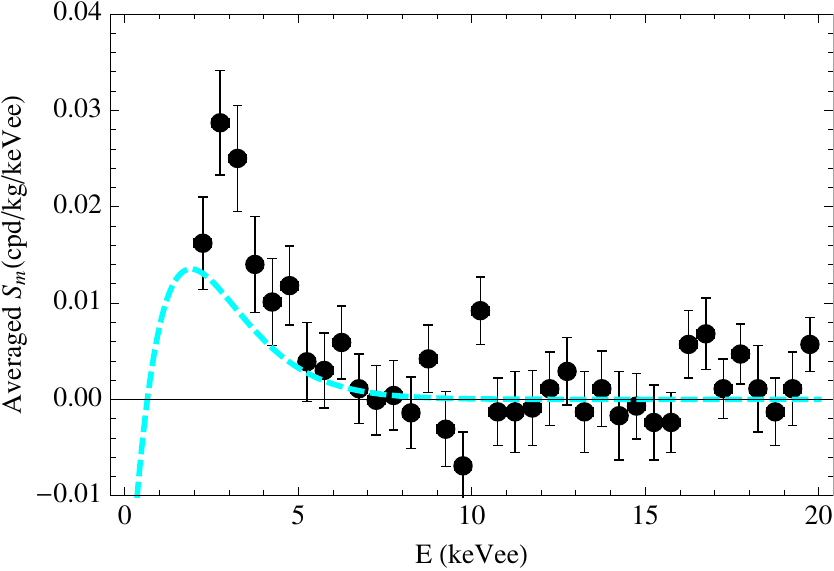}
\end{minipage}
\hspace*{+0.2cm}
\begin{minipage}[t]{0.5\textwidth}
\centering
\includegraphics[width=\columnwidth]{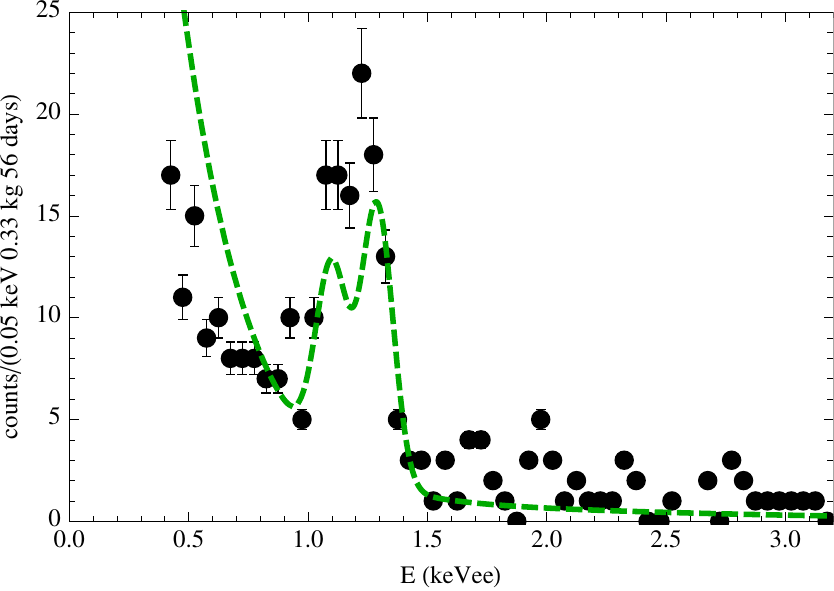}
\end{minipage}
\caption{Same as figure~\ref{fig:fitbyeye}, but for the best-fit point of the extended combined fit, with $m_{\rm DM}$ = 7.38~GeV, $\sigma_n^{\rm SI}=9.64 \times 10^{-41}\ {\rm cm}^2$, 
and other parameter values presented in table~\ref{tab:aastrocomb1}.\label{fig:fitbyeye1}}
\end{figure}

However, despite this shift in the astrophysical parameters,  the extended combined fit offers only a marginal improvement over the standard combined fit.  This can be seen 
in figure~\ref{fig:fitbyeye1}, where we show the spectral shapes of the expected signals for the individual experiments corresponding to the best-fit point of the extended combined fit.
Comparing with figure~\ref{fig:fitbyeye}, we see that the fit to CoGeNT now shows better agreement to the data at low energies, 
while for DAMA the higher value for $q_{\rm Na}$ now leads to a lower peak in the spectrum, which is further suppressed by the smaller value of $\sigma_n^{\rm SI}$ and the lighter dark matter mass.

\begin{table}[t!]
\caption{Same as table~\ref{tab:aastrocomb}, but 
for an extended prior range for $q_{\rm Na}$ (up to $q_{\rm Na} = 0.6$).\label{tab:aastrocomb1}}
\begin{center}
\lineup
\begin{tabular}{ll }
\br
Parameter & Preferred value \\
\br
$v_0$  & $201^{+35}_{-17} \ {\rm km \ s}^{-1}$ \\
$v_{\rm esc}$ & $541^{+27}_{-15} \  {\rm km \ s}^{-1}$ \\
$\rho_{\odot}$ & $0.36^{+0.2}_{-0.09} \ {\rm GeV \ cm}^{-3}$ \\
$q_{\rm Na} $ &$ 0.59^{+0.01}_{-0.04}$\\
${\cal E}_0$  & $9.4 \pm 1.8$~keV\\
$C$ & $3.1^{+2.9}_{-1.6}$~cpd/kg/keV \\
$G_n$ & $2.2\pm 0.4$~cpd/kg/keV \\
\br
\end{tabular}
\end{center}
\end{table}

\section{Conclusions}\label{sec:concl}

The present status of the direct detection of dark matter is somewhat
ambiguous.  On the one hand, there have been claims of detection of a
low-mass WIMP signal from DAMA and CoGeNT.  On the other hand, the
Xenon100, CDMS, and CDMS-II experiments have only been able to provide
exclusion limits.  The interpretation of these experiments in terms of
a dark matter signal is complicated by the presence of backgrounds,
and the need to model experiment-specific systematic effects, such as
the quenching factors for DAMA or the scintillation efficiency for the
Xenon detector.  In addition, one requires the input of astrophysical
quantities, which enter into the theoretical expressions for the total
or modulated DM rates.  All these effects are to some extent subject
to uncertainties, which need to be propagated to the inferred dark
matter parameters.

This multi-parameter inference problem can be addressed in a simple
and consistent way using Bayesian statistical methods.  In the present
work, we apply these methods to a selection of current direct
dark matter searches to infer the mass and cross-section of WIMP dark
matter in the simplest scenario of spin-independent elastic WIMP
scattering.

We initially ignore the astrophysical uncertainties and focus on the
effects of experimental nuisance parameters and background
uncertainties.  Our main result is that the Xenon100 exclusion bound
is significantly weakened once the uncertainty on the scintillation
efficiency is taken into account.  As a consequence, we find that the
CoGeNT preferred region in the $\{m_{\rm DM},\sigma_n^{\rm SI}\}$-plane is
quite compatible with Xenon100, and there is even a marginal
consistency (at $90_S$\% credibility) with the DAMA preferred region.
We expect that this conclusion also holds for Xenon10 and the latest
CoGeNT data which were obtained after two years of data
taking~\cite{cogentAM}.  We also remark that after marginalising over
the background, the standard CDMSGe analysis yields a closed $90_S$\%-credible
contour, although no closed regions remain in the low energy re-analysis CDMSGe(LE).

We then repeat the analysis procedure including astrophysical
uncertainties, considering besides the standard model halo three other
spherically symmetric dark matter halo models with isotropic velocity
distributions, whose density profiles are motivated by $N$-body
simulations.  We find that the inferred values of the astrophysical
parameters are independent of the direct detection experiment data,
indicating that their values depend only on the chosen DM density
profile.  With the exception of the isothermal halo, which prefers
significantly higher escape velocities, the different halo
parameterisations lead to similar values of the astrophysical
parameters.  Not unexpectedly, including the astrophysical
uncertainties further reduces the constraining power of the data in
the $\{m_{\rm DM},\sigma_n^{\rm SI}\}$-plane.  At first glance, this
seems to improve the compatibility between DAMA and CoGeNT.  However,
this impression is somewhat misleading, since in some of the
marginalised directions ($v_0$ and $\rho_\odot$) the disagreement
remains---tweaking astrophysics parameters alone cannot reconcile the
two results. If we demand compatibility between these experiments,
then the inference process naturally concludes that a high value for
the sodium quenching factor for DAMA is preferred.

It will be interesting to apply the analysis framework presented in
this paper to more complex models of the dark matter halo, like
asymmetric velocity distributions or the presence of streams in the
Galactic halo.  Additionally, an application to alternative scenarios
for the particle physics interactions can be envisaged, such as
inelastic DM~\cite{Chang:2010yk,Frandsen:2011ts} or more exotic
scenarios, as for instance discussed
in~\cite{Kopp:2009qt,Chang:2010yk,Graham:2010ca}.

\ack 

We thank Roberto Trotta for useful comments on the manuscript.
CA acknowledges use of the cosmo computing resources at CP3 of
Louvain University. JH gratefully acknowledges support from the
Humboldt foundation via a Feodor Lynen-fellowship.

\appendix

\section{Dark matter density profile in terms of $M_{\rm vir}$ and $c_{\rm cir}$}\label{app1}

The two-parameter DM density profiles defined in section~\ref{sec:haloprofiles}  in terms of $\rho_s$ and $r_s$
can be expressed in terms of the halo's virial mass $M_{\rm vir}$ and concentration parameter $c_{\rm vir}$.   Firstly, the parameter $r_s$ can be parameterised as
\begin{equation}
r_s (M_{\rm vir},c_{\rm vir})= \frac{r_{\rm vir}(M_{\rm vir})}{c_{\rm vir}},
\end{equation}
where the virial radius $r_{\rm vir}$ defines
a spherical region in which the average DM density is $\delta_c=200$ times the critical density $\rho_{\rm crit}$.  The mass enclosed in this region is  called
the virial mass,
\begin{equation}
M_{\rm vir} = 4 \pi \int_{0}^{r_{\rm vir}} \rmd r \ r^2 \rho_{\rm DM}(r) = \frac{4}{3} \pi r_{\rm vir}^3 \delta_c \rho_{\rm crit}\,.
\label{eq:mvir}
\end{equation}
Using this relation, we can solve for $\rho_s$ once a profile has been specified.  We give the solutions for the four halo profiles
considered in this work:

\begin{enumerate}
\item {\it Cored isothermal} :
\begin{equation}
\rho_s (c_{\rm vir})  = \frac{\delta_c \rho_{\rm crit}}{3} \frac{c_{\rm vir}^3}{c_{\rm vir} - {\tan}^{-1}(c_{\rm vir})}\,,
\label{eq:IsoC}
\end{equation}

\item {\it NFW}:
\begin{equation}
\rho_s (c_{\rm vir})  = \frac{\delta_c \rho_{\rm crit}}{3} \frac{c_{\rm vir}^3}{\ln(1+c_{\rm vir})-c_{\rm vir}/(1+c_{\rm vir})}\,.
\label{eq:NFWC}
\end{equation}

\item {\it Einasto}:
\begin{equation}
\rho_s (c_{\rm vir})  = \frac{\delta_c \rho_{\rm crit}}{3} \frac{c^3_{\rm vir} [2^{-\frac{3}{\alpha}} \exp(\frac{2}{\alpha})\alpha^{\frac{3}{\alpha}-1}]^{-1}}{\Gamma\left(\frac{3}{\alpha}\right)-\Gamma\left(\frac{3}{\alpha},\frac{2 c_{\rm vir}^\alpha}{\alpha}\right)} \,,
\label{eq:EinC}
\end{equation}
where $\Gamma(a)$ and $\Gamma(a,b)$ are the gamma and the incomplete gamma functions, respectively.

\item{\it Burkert}:
\begin{equation}
\rho_s (c_{\rm vir})  = \frac{4 \delta_c \rho_{\rm crit}}{3} \frac{c^3_{\rm vir}}{2 \ln(1 + c_{\rm vir}) + \ln(1 + c^2_{\rm vir}) - 2 \tan^{-1}(c_{\rm vir})}\,.
\label{eq:BurC}
\end{equation}

\end{enumerate}

\section*{References}
\bibliographystyle{iopart-num}
\bibliography{biblio}

\end{document}